\documentclass[a4paper,fleqn]{cas-dc}

\usepackage[authoryear]{natbib}
\usepackage{float}
\usepackage{supertabular}

\def\tsc#1{\csdef{#1}{\textsc{\lowercase{#1}}\xspace}}
\tsc{WGM}
\tsc{QE}
\tsc{EP}
\tsc{PMS}
\tsc{BEC}
\tsc{DE}
\begin{document}
\emergencystretch 3em
\let\WriteBookmarks\relax
\def\floatpagepagefraction{1}
\def\textpagefraction{.001}
\shorttitle{}
\shortauthors{Guangzhi Sun et~al.}

\title [mode = title]{Combination of Deep Speaker Embeddings for Diarisation}                      
% \tnotemark[1,2]

% \tnotetext[1]{This document is the results of the research
%   project funded by the National Science Foundation.}

% \tnotetext[2]{The second title footnote which is a longer text matter
%   to fill through the whole text width and overflow into
%   another line in the footnotes area of the first page.}

\author{Guangzhi Sun}
\fnmark[1,2]
\ead{gs534@eng.cam.ac.uk}

\address{Cambridge University Engineering Department, Trumpington Street, Cambridge, CB2 1PZ U.K.}

\author{Chao Zhang}
\fnmark[2]
\ead{cz277@eng.cam.ac.uk}

\author{Philip C. Woodland}
\ead{pcw@eng.cam.ac.uk}
\cormark[1]

\fntext[fn1]{Supported by a Cambridge International Scholarship from the Cambridge Commonwealth, European \& International Trust}
\fntext[fn2]{Equal contributions.}
\cortext[cor1]{Corresponding author.}

\begin{abstract}
Significant progress has recently been made in speaker diarisation after the introduction of \textit{d-vectors} as speaker embeddings extracted from neural network (NN) speaker classifiers for clustering speech segments.
To extract better-performing and more robust speaker embeddings, this paper proposes a \textit{c-vector} method by combining multiple sets of complementary d-vectors derived from systems with different NN components. 
Three structures are used to implement the c-vectors, namely 2D self-attentive, gated additive, and bilinear pooling structures, relying on attention mechanisms, a gating mechanism, and a low-rank bilinear pooling mechanism respectively. 
Furthermore, a neural-based single-pass speaker diarisation pipeline is also proposed in this paper, which uses NNs to achieve voice activity detection, speaker change point detection, and speaker embedding extraction. Experiments and detailed analyses are conducted on the challenging AMI and NIST RT05 datasets which consist of real meetings with 4--10 speakers and a wide range of acoustic conditions. 
For systems trained on the AMI training set, relative relative speaker error rate (SER) reductions of 13\% and 29\% are obtained by using c-vectors instead of d-vectors on the AMI dev and eval sets respectively, and a relative SER reduction of 15\% in SER is observed on RT05, which shows the robustness of the proposed methods. By incorporating VoxCeleb data into the training set, the best c-vector system achieved 7\%, 17\% and 16\% relative SER reduction compared to the d-vector on the AMI dev, eval and RT05 sets respectively.

\end{abstract}

% \includegraphics{grabs.pdf}

\begin{highlights}
\item Combining deep speaker embedding (window-level d-vector) extraction systems using 2D self-attentive, gated additive and bilinear pooling methods. The best-performing  structure for combination is obtained by stacking a 2D self-attentive and a bilinear pooling structures.
\item A complete single pass neural network-based diarisation pipeline is introduced, which includes neural voice activity detection, neural change point detection, a deep speaker embedding extraction system and spectral clustering.
\item Experiments on both the AMI and the NIST RT05 evaluation sets showed that our proposed methods can produce state-of-the-art results for the very challenging multi-speaker (with 4--10 speakers) meeting diarisation task.
\item We use the AMI dataset based on the official speech recognition partition with the audio recorded by multiple distance microphones (MDM) since this is a more realistic setup for meeting transcription than many different setups used by previous AMI based studies. 
% This is the first paper to use this setup to the best of the authors' knowledge. 
Although this makes our results not directly comparable to those from the previous papers, we think our system still shows superior performance, 
since the results shown in Table~\ref{tab:finalresult} are the lowest diarisation error rates with the same training data, and the realistic setup we used can increase the difficulty of the task. 
With this set-up, our best combination method achieved 13\%, 29\% and 15\% relative SER reductions on the AMI dev, eval and RT05 sets respectively compared to the better individual system.

\item Further reductions to all systems in DER were found using VoxCeleb and VoxCeleb2 data for training. Moreover, using extra training data, our best combination method achieved 7\%, 17\% and 16\% relative SER reductions on the AMI dev, eval and RT05 sets respectively compared to the best individual system.

\item This paper significantly extends work in on our previous conference paper \citep{mypaper}. Compared to our previous paper
\begin{itemize}
    \item Almost all of the combination structures are newly proposed and not included in \citep{mypaper}, apart from the first type of 2D self-attentive method defined in Eqn.~\eqref{eq:consec1}.
    \item \citet{mypaper} only performed the experiments on the AMI data with manual segmentation, while this paper explores to use both manual and automatic segmentation. The neural diarisation pipeline is newly introduced in this paper, although the neural VAD structure was  proposed in \citep{vad}. 
    \item None of the results given in this paper, for both AMI and RT05, appeared in \citep{mypaper} or \citep{vad}. The NIST RT05 evaluation set was not used in our previous papers and confirms our main results on another standard task.
\end{itemize}

%proposed combination methods outperform individual systems by clear margins.

\end{highlights}


\maketitle \footnote{\textcopyright 2021 This manuscript version is made available under the CC-BY-NC-ND 4.0 license \url{http://creativecommons.org/licenses/by-nc-nd/4.0/}}
\section{Introduction}
Multi-party interactions such as meetings and conversations is one of the most important tasks for many speech and language applications. Speaker diarisation, the task of finding ``who spoke when'' in a multi-speaker audio stream, is critical for such applications, and has received increasing attention in recent years. Speaker diarisation typically involves splitting the audio stream into many speaker homogeneous  segments based on the detected voice activity and speaker change points, and clustering those segments into groups that  correspond to the same speaker. Recent systems often implement the clustering step by first converting each variable-length segment into a fixed-length vector representing the speaker identity, referred to as a \textit{speaker embedding}, and then performing clustering based on these vectors. Speaker embedding are also widely used in other spoken language processing tasks, such as speaker recognition, speech recognition, and speech synthesis \citep{speakeremb1,speakeremb2,speakeremb3}.

Traditionally, \textit{i-vectors} obtained using Gaussian mixture models \citep{ivec2} are the most widely used speaker embeddings that have been applied to speaker diarisation \citep{ivec1,ivec3,ivec4,ivec5}.
With the development of deep learning, \textit{d-vectors} \citep{dnn1} have been proposed as deep speaker embeddings  derived as the output from a hidden layer of a neural network (NN) for classifying the training set speakers at the frame-level. It has been shown that d-vectors can outperform i-vectors in both speaker recognition \citep{dnn1,dnn2} and speaker diarisation \citep{diardeep2}. %Apart from feed-forward models with fully connected (FC) layers, recurrent and convolutional models are also used to extract d-vectors at frame-level \citep{rcnn1,rnn1}. 
By using a pooling function across time in an input window covering for instance 200 frames, d-vectors can be extracted at the window level \citep{dnn3,xvec,selfatten1}.
In addition, NNs can be applied to other diarisation components, such as voice activity detection (VAD) \citep{vad0,vad1,vad} and speaker change point detection (CPD) \citep{dnncpd}.
Recently, NNs have also been used to replace the clustering algorithm \citep{e2e3,dnc,gnnsc}.

%various diarisation systems have integrated deep neural networks (DNN) into different parts of the pipeline \citep{dnncpd,diardeep2,dnn3}. One of the most significant contribution from deep learning is the introduction of d-vectors which are speaker embeddings based on NNs trained for speaker classification. D-vectors have shown significant improvements over the traditional i-vectors based on factor analysis \citep{ivec1,ivec2,ivec3,ivec4,ivec5}. A wide range of neural network architectures have been applied for d-vector extraction, including deep feed-forward models \citep{dnn1,dnn2,dnn3}, recurrent neural networks (RNN) \citep{rnn1} and convolutional neural networks \citep{rcnn1}. 

%Although the use of various deep learning structures has achieved improvements, d-vectors extracted using a certain network structure have their strengths and weaknesses. 
Although the use of more powerful NNs often leads to more expressive d-vectors, different model structures have different innate strengths and weaknesses. Therefore, it is desiarble 
%to exploit the complementarity among different d-vector systems and 
to derive better speaker embeddings by combining d-vectors extracted using different NN systems \citep{mypaper}.
%effectively based on their complementarity .
%Therefore, effective combinations to take advantage of the complementarity among different d-vector systems is desired. %In this paper, we extend our previous work on using a combination  \citep{mypaper} 
In this paper, the \textit{c-vector} approach which refers to the combination of deep speaker embeddings is proposed. 
%Based on the generic combination framework proposed by \citet{mypaper}, three different combination methods are explored, namely the 2D self-attentive combination, the gated additive combination and the bi-linear pooling combination. The 2D self-attentive combination extends the idea of d-vector extraction using temporal attention \citep{selfatten1} to attentively combine speaker embeddings across time and systems consecutively.
Three different methods are explored for d-vector combination: namely 2D self-attentive; gated additive; and bilinear pooling structures. Combination structures can be jointly optimised with all d-vector extraction systems to be combined. The multi-head self-attentive structure \citep{selfattent0} is used as the temporal pooling function to generate d-vectors for all systems at the window-level throughout this paper. If a second self-attentive structure is applied to integrate d-vectors from different systems, the combination method becomes a 2-dimensional (2D) self-attentive one. 
%Each d-vector is entirely scaled by the attention weight which is determined by the d-vector itself to indicate its importance. 
%Gated additive structure also combines systems by summation, whereas the scaling factor is different for different element in each d-vector via a gating mechanism \citep{lstm}.
The gated additive structure uses the gating mechanism \citep{lstm} which sums d-vectors whose individual elements are scaled with different and dynamic scaling factors.
% where a weighted sum is used to combine the d-vectors and the scaling factors are dynamic and different for each element in each d-vector.
Bilinear pooling, on the other hand, integrates d-vectors based on the outer product \citep{bilinear_origin}, where a multiplication of every pair of elements from two vectors is calculated for the combination. Bilinear pooling provides rich factor interactions between speaker embedding models in a multiplicative way in contrast to the aforementioned combination methods. The bilinear pooling structure in this paper is based on the low-rank bilinear pooling method \citep{bilinear}, which uses a low-rank approximation and the Hadamard product to reduce the number of parameters and calculations. Moreover, the low-rank bilinear pooling technique is modified by adding residual connections,
%in an alternative way, 
which improves the training stability and the resulted c-vector performance. 
%systems in a multiplicative way using the Hadamard product based on the low-rank bi-linear pooling method \citep{bilinear}. Residual connections are also added to the bi-linear pooling combination for stable training and better performance. Combined speaker embeddings are referred to as "c-vectors" throughout the paper.

To show the effectiveness of c-vectors, both time-delay neural networks (TDNNs) \citep{tdnn0,tdnn} and high order recurrent neural networks (HORNNs) \citep{hornn} are selected as example feed-forward and recurrent NN structures to generate complementary d-vectors for combination. In addition to combination methods, a neural-based single pass diarisation pipeline is also proposed in this paper which uses NNs for VAD, CPD, and speaker embedding extraction. %Apart from clustering, the other three stages are built with networks trained on classification tasks.
%Spectral clustering \citep{speccluster,sc_dia1,ivec3,rnn1} is used to cluster the speaker embeddings, and other similar methods, such as agglomerative clustering \citep{iac_dia1,iac_dia2,ivec1} and neural clustering \citep{dnc,gnnsc}, can also be used in this pipeline.
Spectral clustering is used \citep{speccluster,sc_dia1,ivec3} to cluster the speaker embeddings throughout the paper, 
%, and other similar methods, such as agglomerative clustering \citep{iac_dia1,iac_dia2,ivec1} can also  be used in this pipeline. 
which can be extended to a full neural pipeline if neural clustering \citep{dnc,gnnsc} is used. 

The remainder of this paper is organised as follows. Section \ref{sec:relatedworks} reviews related work. Section \ref{sec:pipeline} introduces the proposed speaker diarisation pipeline and Section \ref{sec:model} describes the combination methods in detail. 
Sections \ref{sec:expsetup} and \ref{sec:results} present the experimental setup and results. Finally, the conclusions are given in Section \ref{sec:conclusions}.

\section{Related work}
\label{sec:relatedworks}
%Works on i-vector and d-vector combination
Speaker diarisation %\citep{review1,review2} 
is a long-standing research topic and is often applied to data with long audio streams, such as telephone conversations, meetings \citep{limsi_meeting}, and broadcasts \citep{limsi_broadcast}. \citep{review2} provides an excellent review of the research problem and early work on speaker diarisation.
%It had been referred to as speech segmentation and clustering before the term ``diarisation'' was first introduced in the DARPA EARS project proposal in 2002.
Although many 
%various implementations of 
speaker diarisation systems have been developed, most pipelines start with a VAD component
% Although how to implement a speaker diarisation system is still an open problem, most  pipelines start with a VAD component,
which finds speech segments in the audio stream. 
%used to employ a phone recogniser or apply some energy constraints \citep{vad2,vad3}, while recently some systems started to adopt neural networks \citep{vad}.
Traditional VADs use the zero-crossing rate, energy constraints, or a phone recogniser \citep{zcrvad,vad2,vad3}, %used to employ a phone recogniser or apply some energy constraints \citep{vad2,vad3}, while recently some systems started to adopt neural networks \citep{vad}.
while more recent systems have used NNs to classify speech and non-speech \citep{vad0,vad1,vad}.

After obtaining regions of audio containing speech, CPD can be applied to divide the initial speech segments into speaker homogeneous segments. CPD methods often fall into three categories: model-based methods, distance-based methods, and hybrid methods \citep{cpd1,cpd2,cpd3}. Recent studies focus on model-based CPD using NNs with different structures, such as feed-forward \citep{dnncpd}, convolutional \citep{cnncpd}, and recurrent \citep{rnncpd} models. 
For clustering speech segments, some early studies used the Bayesian information criterion for bottom-up clustering \citep{bottomup1,bottomup2,bottomup3,iac_dia2}, while the others use top-down approaches using hidden Markov models \citep{topdown1,topdown3} for joint optimisation of segmentation and clustering.

%some early researches have adopted Bayesian information criterion approaches which involves a CPD stage followed by a bottom-up clustering \citep{bottomup1,bottomup2,bottomup3}, while others tried to use top-down approaches involving hidden Markov models (HMM) which enables a joint optimisation of segmentation and clustering \citep{topdown1,topdown2,topdown3}. CPD, also known as speaker segmentation, usually fall into three categories including model-based methods, distance-based methods and hybrid methods \citep{cpd1,cpd2,cpd3}. In particular, recent advances in model-based methods have employed a range of trained NN models including DNN \citep{dnncpd}, CNN \citep{cnncpd} and RNN \citep{rnncpd}.

When clustering speaker homogeneous segments into a number of speaker clusters, it is also convenient to first convert each variable-length segment into a fixed-length vector representation, which allows common clustering algorithms, such as $k$-means, spectral clustering \citep{sc_dia1}, and agglomerative clustering \citep{vad2,iac_dia2} to be used. 
Although early studies on speaker adaptation with cluster adaptive training and eigenvoices \citep{ivec0} explored vector speaker representations, the i-vector method which is based on joint factor analysis in the total variability space, became the most widely used before deep learning \citep{ivec2}. Clustering i-vectors based on the cosine distance has been widely adopted in diarisation systems \citep{ivec1,ivec4,ivec6,meanshiftcluster}.
% Using cosine distance as the metric, i-vectors have been widely used in diarisation with different clustering algorithms \citep{ivec1,ivec4,ivec6,meanshiftcluster}.
Furthermore, variational Bayes methods are often used to refine the i-vector-based clustering results or segment boundaries \citep{vbivec,pldacluster,ivec3,ivec5}.

With the advent of deep learning, d-vectors \citep{dnn1} 
%which are output vectors derived from a hidden layer of a DNN-based training set speaker classifier, 
have been used to replace i-vectors. 
In addition to feed-forward deep neural networks (DNNs) \citep{dnnam}, recurrent and convolutional models have also been applied to extract d-vectors at the frame-level \citep{dnn1,dnn4,dnn2,rcnn1,rnn1}. To convert a variable length segment into a fixed-length vector using frame-level d-vectors, a temporal pooling function, such as the mean and standard deviation \citep{dnn3,tdnn1,selfatten2}, attention mechanisms \citep{dnn5,selfatten1,mypaper,selfatten3}, and their combination \citep{statspool1}, have been used, which also enables joint training over entire segments.

% A temporal pooling function, such as the mean and standard deviation \citep{dnn3,tdnn1,selfatten2}, attention mechanisms \citep{dnn5,selfatten1,mypaper,selfatten3}, and their combination \citep{statspool1}, is important to achieve not only the conversion from a variable-length speech segment into a fixed-length d-vector, but also the joint training over the entire segment.

To address the mismatch between training and test, innovative loss functions, such as end-to-end losses \citep{dnn2,triplet2,tdnn1} and angular softmax losses \citep{softmax4,softmax5,softmax6,asoftmax,softmax7,losscomb,asoftmax2} has been used to train speaker embeddings. In some recent research, clustering can either be merged with prior stages for joint optimisation \citep{e2e1,e2e3,e2e4,e2e5,e2e6}, or combined with discriminative training \citep{dnc}.

System combination for speaker diarisation has been actively studied, including system fusion at either the output score level \citep{fuse3,fuse1,fuse2,veccomb3} or the speaker embedding level \citep{fcfuse,veccomb}. Specifically, \citet{fcfuse} and \citet{veccomb} fuse an i-vector with a d-vector using a fully-connected (FC) layer and an attention mechanism respectively. \citet{veccomb3} averages scores from two systems with different NN structures for d-vector extraction. \cite{speakerGAN} concatenates continuous and discrete latent variables from a generative adversarial network.
 %Research studies on system combination for speaker diarization in recent years have involved fusion of systems with different sources \citep{fuse1}, different speaker embeddings \citep{fcfuse,veccomb}, different training objectives \cite{fuse2} and different outputs \citep{fuse3}. In particular, \citet{veccomb} proposes a framework that learns a joint representation from i-vectors and RNN-based d-vectors, and \citet{veccomb3} combines two different neural architectures for speaker characterisation.

% Sec. 3: Pipeline
\section{Speaker diarisation pipeline}
\label{sec:pipeline}
\begin{figure*}[width=2.0\linewidth, pos=t]
    \centering
    \includegraphics[scale=0.32]{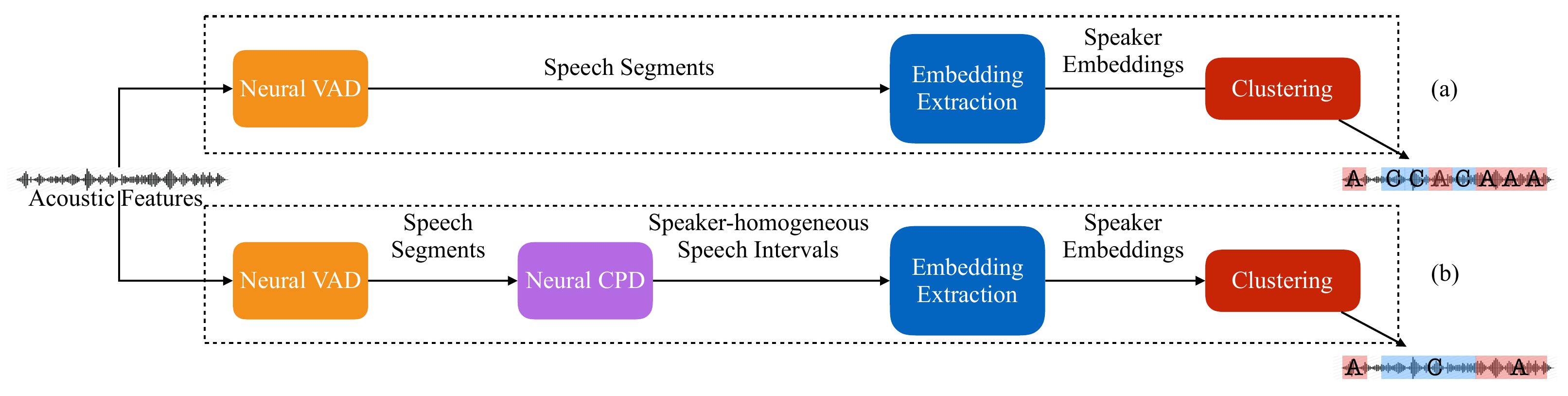}
    \vspace{-0.5cm}
    \caption{Illustration of the speaker diarisation system. Path (a) denotes the window-level clustering pipeline. Path (b) denotes the full pipeline with the CPD component, which leads to the speaker-homogeneous segment-level clustering results.}
    \vspace{-0.5cm}
    \label{fig:pipeline}
\end{figure*}

The diarisation pipeline used in this paper is shown in Fig. \ref{fig:pipeline} which has four components: VAD, CPD, speaker embedding extraction, and clustering. 
Given an input audio stream, VAD detects audio segments containing speech which are then
%distinguishes between its speech and non-speech parts and splits them into segments. Segments corresponding to speech are 
further processed by the CPD component and split into speaker homogeneous segments. 
A speaker embedding is extracted for each of these segments using an NN model. Finally, a spectral clustering algorithm is used to group similar speaker embeddings and assign one speaker label to all segments within each group. 
In this paper, all components, except for the clustering algorithm based on cosine similarity, use neural networks. Alternatively, if each speech segment from VAD is split into multiple windows with the same length, speaker clustering and label assignment can be performed directly on such windows. This is referred to as window-level clustering which bypasses the CPD stage as shown in Fig. \ref{fig:pipeline}.

%VAD distinguishes between speech and non-speech parts in the given audio stream, and selects segments corresponding to speech parts to be sent to the CPD. CPD performs a further segmentation on the VAD output to generate speaker homogeneous segments. Speaker embeddings are then extracted for each segment using a neural network model trained on speaker classification tasks. Finally, the clustering algorithm groups similar embeddings together and assigns a speaker to each segment based on the group it belongs to, which leads to the final diarisation result. Except for the clustering stage which uses cosine similarity-based spectral clustering algorithm on speaker embeddings, each stage is built based on a neural network model. Alternatively, speech segments from VAD can be split into fixed-length segments, known as "windows", and clustering and speaker assignments are performed directly on those windows. This is referred to as the window-level clustering which bypasses the CPD stage as shown in Fig. \ref{fig:pipeline}.

\subsection{Neural VAD and CPD}
\label{ssec:vadcpd}
\begin{figure}[h]
    \centering
    \includegraphics[scale=0.4]{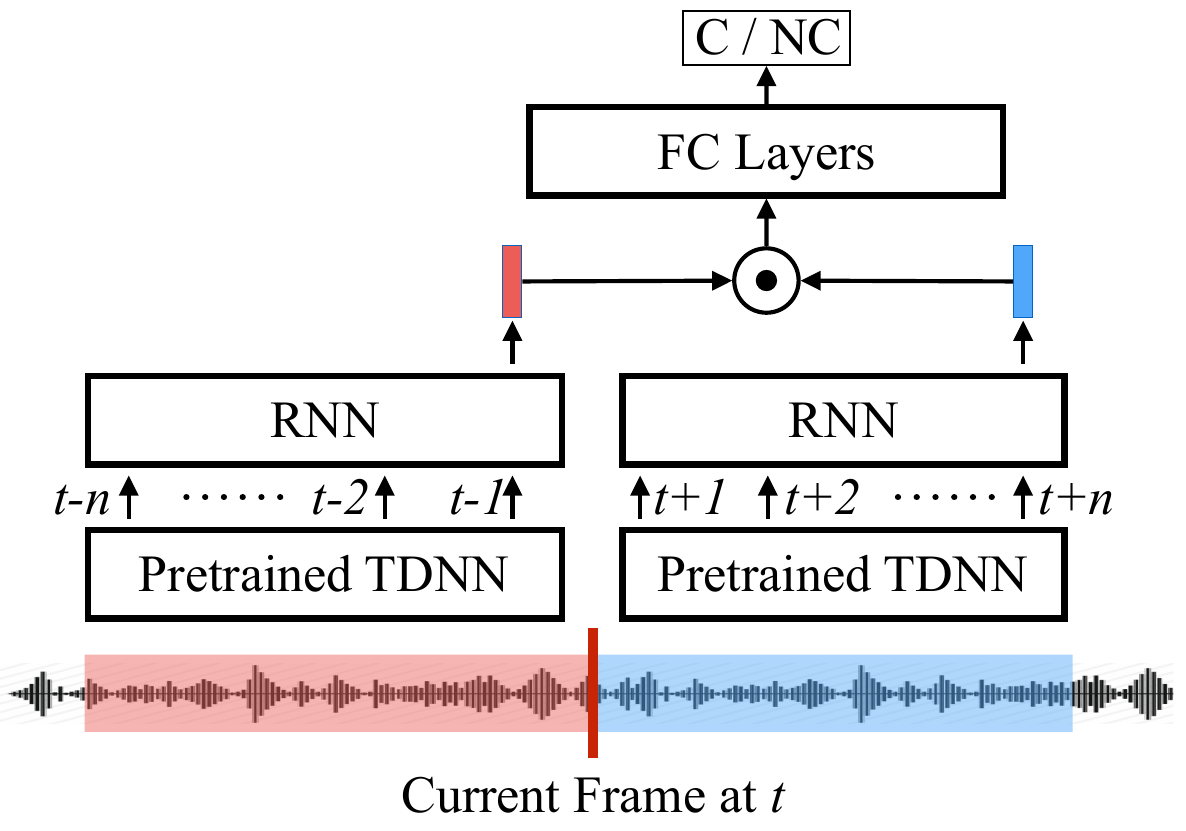}
    \caption{Proposed neural CPD structure to distinguish speaker change (C) and non-change (NC) points. The TDNN and the RNN on both sides of the current time share the same parameters. $\odot$ represents the Hadamard product.}
    \label{fig:cpd}
\end{figure}

The VAD and CPD models are both built as NN-based frame-level binary classifiers and described below:
%based on 40-dimensional (-d) log-Mel filter bank (FBK) features, whose detailed structures are:
%Both VAD and CPD are built as NN-based frame-level binary classifiers: the VAD model performs the speech and non-speech classification and the CPD model performs the speaker change and non-change classification.
%, where the neural network predicts posteriors for hidden Markov model (HMM) states and frame-wise decisions are made using a Viterbi decoder \citep{vad}. 
%The detailed structures of the two models are:
\begin{itemize}
    \item The VAD model is a DNN which consists of seven FC layers with ReLU activation functions. The key strength of the DNN structure is the use of a large input window covering 55 consecutive frames (27 on each side), which provides sufficient information for high performance speech and non-speech classification \citep{vad}.

    \item The CPD model, as shown in Fig.~\ref{fig:cpd},
    % has a complex structure. Our CPD model 
    uses a ReLU recurrent neural network (RNN) model to encode past and future input into two separate vectors respectively which are then fused into one vector using the Hadamard product followed by a softmax FC layer classifying speaker change or non-change. By viewing the RNN output vectors as speaker representations corresponding to the past and future audio segments, the Hadamard product and the output layer can be seen as making decisions on the change of speaker identity by comparing the speaker representations before and after the current time.    
    
    To further improve the ability of the RNN to extract speaker representations, speaker embeddings instead of raw acoustic features are used as the input to the RNN model. As shown in Fig.~\ref{fig:cpd}, a TDNN trained to classify the training set speakers is used to extract the frame-level d-vectors which are fed into the RNN. 
    %To avoid the local optimum issue caused by pre-training, 
    The whole CPD model including the TDNN, the RNN, and the output layer are then jointly trained to perform speaker change/non-change classification. 
    
    %A ReLU TDNN is first pretrained to perform training set speaker classification at frame-level \citep{tdnn}. The first FC layer of the TDNN converts each input vector with 5 consecutive frames (2 on both sides) into a 256-d vector, and a concatenation of such vectors at -2, 0, and +2 (the second time step to the current time, the current time, and the second time step in the future) is transformed by the second FC layer into another 256-d vector. Similarly, the third FC layer converts the concatenation of the output vectors from the second FC layer at -3, 0, +3 into the third 256-d vector, which is first transformed by the fourth FC layer into 128-d and then fed into the softmax output layer. Overall, this TDNN structure covers an input window with 15 consecutive frames (7 on both sides). Afterwards, the output vectors derived from the penultimate layer of the TDNN speaker classifier, which are actually frame-level d-vectors, are used as the initial input values to the rest of the CPD model. 
    
%Next, the d-vectors derived from the TDNN at -42, -28, -14, 0, are then encoded into a longer-term past time vector by a ReLU recurrent neural network (RNN) layer. 
    
%that a TDNN model is first pretrained to perform speaker classification at frame-level, to provide discriminative features towards speaker identities. A recurrent neural network (RNN) with ReLU activation function is used to encode 
\end{itemize}

. %The system is trained at frame-level with cross-entropy criterion. 
%\begin{figure}[h]
%    \centering
%    \includegraphics[scale=0.39]{VAD.pdf}
%    \caption{A sketch map of our neural VAD. The left and right examples predict their current frames as speech (S) and non-speech (NS) respectively.}
%    \label{fig:vad}
%\end{figure}
%Similarly, CPD is trained on change/non-change classification at frame-level, with a more complicated structure that compares the left and write segments about the current frame, as shown in Fig \ref{fig:cpd}.
%The TDNN is first pre-trained on speaker classification to generate speaker-discriminative features for later comparison. To avoid local inconsistency of a single speaker begin captured, an RNN encoder is applied across a sequence of frame-level TDNN output to cover a wide range of frames. The same network with same parameters is applied on both sides of the current frame, and outputs from both sides will be compared using Hadamard product followed by fully-connected layers with a binary classification output as in VAD. 

After VAD, the original input audio stream is split into many variable length segments that may or may not contain speech, and each speech segment may have multiple speakers. Such speech segments can be directly used for speaker clustering, as shown by the dotted line in Fig.~\ref{fig:pipeline}, or further split by the CPD model into speaker-homogeneous segments before the speaker embedding extraction and clustering stages. 

\subsection{Speaker embedding extraction}
%Speaker embeddings, including d-vectors and c-vectors proposed in the next section, are extracted from a deep neural network trained on speaker classification task. During training and test, variable length segments are first split into fixed-length segments, referred to as "windows", with certain overlaps. Then, speaker embeddings are extracted for each window. These window-level speaker embeddings will be sent to the Softmax output layer for classification during training, and will be dumped out for clustering during test.
In this paper, Speaker embeddings, including both d-vectors and the proposed c-vectors, are the penultimate layer outputs of an NN trained to perform classification among training set speakers. 
In both training and test, each variable length speech segment  %(which is considered as speaker-homogeneous if CPD is used in Fig.~\ref{fig:pipeline}) 
is first split into multiple fixed-length windows with a certain amount of overlap, and a speaker embedding is then extracted from each window. At test-time, the speaker embeddings are used as input vectors to the clustering algorithm. The detailed model structures for speaker embedding extraction are presented later in Section~\ref{sec:model}.

\subsection{Clustering}
\begin{figure}[h]
    \centering
    \includegraphics[scale=0.3]{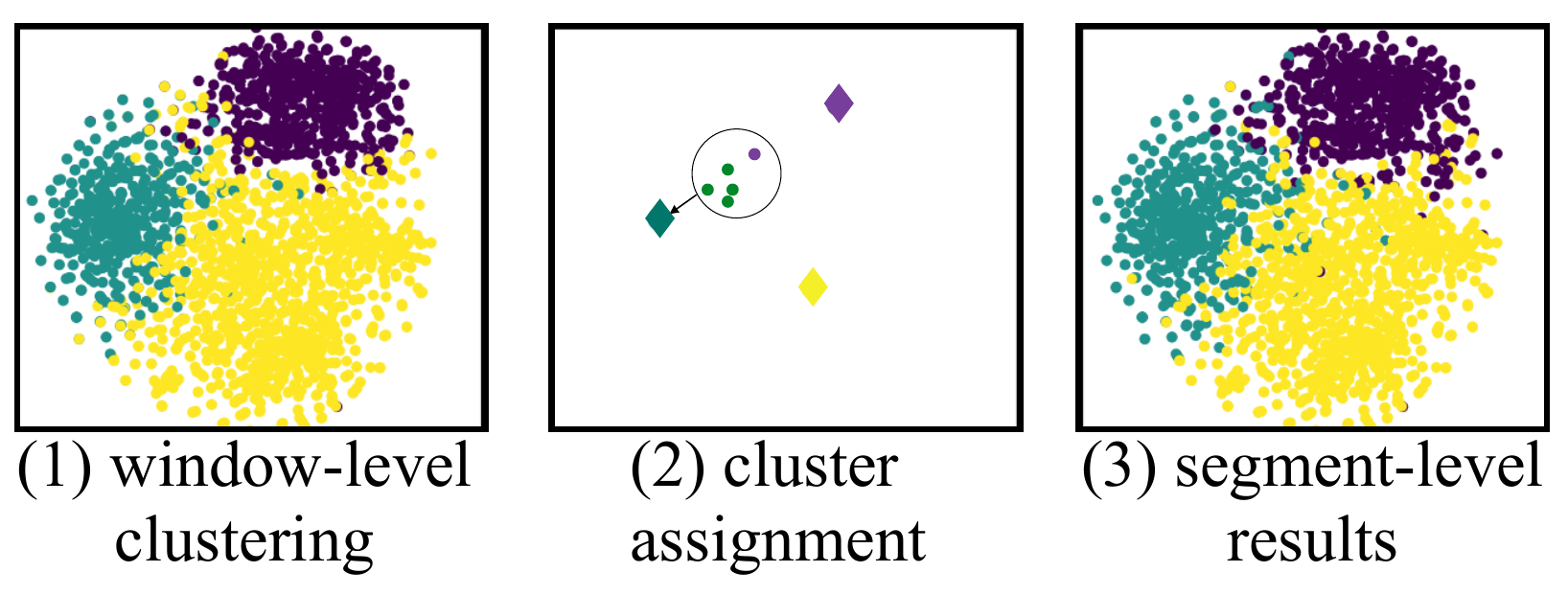}
    \vspace{-0.5cm}
    \caption{t-SNE 2D plots showing the clustering process of a meeting TS3004d from the AMI corpus. Each dot represents one speaker embedding extracted from a 2 second window. The three diamonds represent centroids of three clusters. Sub-figure (1) shows the window-level clustering results based on speaker embeddings, which corresponds to the results produced by Path (a) in Fig. \ref{fig:pipeline}. Performing the steps in sub-figures (1), (2) and (3) sequentially corresponds to the process in Path (b) in Fig. \ref{fig:pipeline}.
    Sub-figure (2) shows the procedure for cluster assignment where dots in the circle belong to the same speaker-homogeneous segment and the arrow points to the assigned cluster. A neural CPD is required to generate the speaker-homogeneous segments. 
    Sub-figure (3) shows the clustering results at the speaker-homogeneous segment-level.}%Path (a) in Fig. \ref{fig:pipeline} only adopts the procedure in sub-figure (1) while path (b) in Fig. \ref{fig:pipeline} uses all three procedures.}
    \label{fig:tsne}
    \vspace{-0.5cm}
\end{figure}

A modified spectral clustering \citep{rnn1} approach based on the cosine distance, together with a post-processing stage, is used to cluster speaker embeddings.
% \begin{figure}[h]
%     \centering
%     \includegraphics[scale=0.38]{clustering.pdf}
%     \caption{Spectral clustering and post-processing stages.}
%     \label{fig:clustering}
% \end{figure}
Spectral clustering is first performed on window-level speaker embeddings where the number of clusters is determined by the maximum eigenvalue drop-off \citep{speccluster}, and is set to be greater than one. 
%Then, if CPD is used and the variable-length speech segments are speaker-homogeneous, the cosine distance between the average of the window-level speaker embeddings of each segment and the centroid of each cluster is calculated, and the cluster whose centroid is  nearest to the average of the speaker embeddings will be assigned to that speech segment. 
Next, if CPD is used, variable-length speech segments after CPD are treated as speaker-homogeneous, and each segment is assigned to a cluster whose centroid has the smallest cosine distance to the average of the window-level speaker embeddings from this segment. 
Otherwise, a window-level speaker-homogeneous assumption is made, and different input windows of the same speech segment are allowed to be assigned to different cluster labels. For visualisation, 2D plots using t-distributed stochastic neighbour embedding (t-SNE) for dimensionality reduction is shown in Fig. \ref{fig:tsne} where sub-figure (1) corresponds to Path (a) in Fig. \ref{fig:pipeline} and and sub-figures (1), (2) and (3) together show the process of Path (b) in Fig. \ref{fig:pipeline}. Boundaries between two clusters are changed due to cluster assignment.

% Sec. 4: Model
\section{Combination of deep speaker embeddings}
\label{sec:model}
%This section first describes the model structures used for segment-level d-vector extraction. Three model-based d-vector combination methods are proposed in this paper, which rely on the self-attentive structure, the gating structure, and the bilinear pooling structure to combine the segment-level d-vectors derived from two different systems respectively. 
%namely 2D self-attentive combination, gated additive combination, and bilinear pooling combination. 
This section describes the three proposed structures to combine the window-level d-vectors for c-vector extraction using an attention mechanism, a gating mechanism, and a bilinear pooling mechanism. Window-level d-vectors are derived from multiple complementary neural network structures to obtain an enhanced speaker embedding with improved performance. 
A multi-head self-attentive structure with a modified penalty term is used as the temporal pooling function to derive the window-level d-vectors. Similar to the extraction procedure for d-vectors, the whole c-vector extraction model, including d-vector extraction and combination, is optimised in an end-to-end fashion to classify the training set speakers. Output vectors of the penultimate network layer are used as the c-vectors for speaker clustering.
% representing the unseen speakers. (dev set has seen speakers)
%It is worth noting that though it is straightforward to apply the proposed methods to combine any number of d-vectors, only the case with two complimentary d-vectors is presented for simplicity and clarity. 

The first structure, 2D self-attentive combination, integrates multiple window-level d-vectors into a c-vector using an extra multi-head self-attentive structure. The second structure, gated additive combination, scales each element of each window-level d-vector individually and dynamically based on the gating mechanism. Compared with the 2D self-attentive combination, the gated additive combination transforms each d-vector in a more flexible way before summing them into a combined vector. Instead of the dynamic weights used in the first two structures,
bilinear pooling combination employs a separate static weight to scale the multiplicative interaction of every pair of elements from the window-level d-vectors, which enables a powerful combination method with a large number of static weights. An improved low-rank approximation with residual connections is proposed and applied to the bilinear pooling combination, in order to reduce complexity and avoid potential over-fitting issues. Lastly, a stacked structure that combines 2D self-attentive combination with bilinear pooling combination is proposed, which further leverages the complementarity of the dynamic-weight-based and static-weight-based structures.
%relies on the attention mechanism to integrate the window-level d-vectors into a c-vector, which assigns a dynamic scalar weight to each d-vector
%leverages the attention mechanism that scales each of the window-level d-vectors with a dynamic scalar weight and sums the scaled vectors together. Multiple sets of dynamic weights can be produced and the summed vectors can be concatenated to form the final c-vector.  
%combines the window-level d-vectors using weighted sums based on multiple sets of dynamic scalar weights produced by an extra attention structure. 
%Different from the attention mechanism that scales each vector with a dynamic scalar weight, 

\subsection{2D self-attentive combination}
\label{ssec:selfatt}
The 2D self-attentive combination method, as its name suggests, extracts c-vectors by combining d-vectors in two dimensions using the multi-head self-attentive structure \citep{selfattent0} with a modified penalty term \citep{mypaper}. One combination dimension integrates the frame-level d-vectors extracted across time into a single window-level d-vector representing the entire window. The other combination dimension is to fuse across the window-level d-vectors produced by multiple systems. %Both consecutive and simultaneous structures for 2D self-attentive combinations were proposed in our previous paper \citep{mypaper}. In particular, this paper investigates the consecutive combination which uses two levels of separate self-attentive layers to first combine across time and then across systems.

\subsubsection{Multi-head self-attentive structure}
\label{sssec:selfattentive}
First, the multi-head self-attentive structure and the modification of the penalty term is presented. 
%The essential component of the 2D self-attentive combination method is the multi-head self-attentive layer, which dynamically calculates a set of weighted averages of input speaker embeddings to be combined. 
As a type of attention mechanism, the self-attentive structure dynamically calculates a set of weights, termed an annotation vector, to integrate the input sequence into one vector through a weighted average. When used to integrate the frame-level d-vectors across time, the dynamic weights  in the annotation vector directly reflect the speaker-discriminative ability of the frame-level d-vectors at different time steps. 
%Multiple annotation vectors can be generated from multiple attention output heads, and more information can be reserved in the window-level d-vector by concatenating the integrated vectors related to each annotation vector. 
To encapsulate diverse speaker characteristics, multiple annotation vectors can be generated from multiple attention output heads to combine the frame-level d-vectors using different sets of dynamic weights.
The structure is ``self-attentive'' since the input sequence used to compute the annotation vectors is simply the sequence of frame-level d-vectors which are to be combined.

Specifically, if $\mathbf{h}(t)$ is a frame-level d-vector at time $t$, $T$ is the length of the input window, the input to the attentive structure is a $T\times N$ matrix $\mathbf{H}  =[\mathbf{h}(1),\mathbf{h}(2),\dots,\mathbf{h}(T)]^{\text T}$, where $N$ is the size of $\mathbf{h}(t)$. Let $\mathbf{A}$ be the $T\times G$ annotation matrix formed by the $G$ annotation vectors, $\mathbf{E}$ be the $G\times N$ output matrix formed by $G$ integrated vectors, $\mathbf{A}$ and $\mathbf{E}$ can be computed by
%can be obtained using Eq.~(\ref{eq:weight}). The integrated vector applied to the inputs as in Eq.~(\ref{eq:output})
%a $T\times n$ matrix $\mathbf{H}  =[\mathbf{h}(1), \mathbf{h}(2), \dots, \mathbf{h}(T)]^T$ where $n$ is the dimension of each vector, the annotation matrix can be calculated using Eq.~(\ref{eq:weight}) and applied to the inputs as in Eq.~(\ref{eq:output})
\begin{align}
\mathbf{A}&=\text{Softmax}(\tanh(\mathbf{H}\mathbf{W_1})\mathbf{W_2}), \label{eq:weight}\\
\mathbf{E}&=\mathbf{A}^{\text T}\mathbf{H},
\label{eq:output}
\end{align}
where $\text{Softmax}(\cdot)$ and $\text{tanh}(\cdot)$ refer to the softmax and hyperbolic tangent activation functions. From Eqn.~\eqref{eq:weight}, the attentive structure is a feedforward neural network model with two FC layers, whose weight matrices are $\mathbf{W_1} \in \mathbb{R}^{N\times N}$ and $\mathbf{W_2} \in \mathbb{R}^{N\times G}$. The input and output matrices are $\mathbf{H}$ and $\mathbf{A}$, and the softmax function is performed over each column of $\mathbf{A}$. We denote the self-attentive structure defined in Eqns.~\eqref{eq:weight} and \eqref{eq:output} as
\begin{align}
\label{eq:selfatten}
\mathbf{E}=\text{SelfAtt}\left(\mathbf{h}(1),\mathbf{h}(2),\ldots,\mathbf{h}(T)\right).
%\mathbf{E}=\text{SelfAtten}(\mathbf{H})=\text{SelfAtten}(\mathbf{h}(1),\mathbf{h}(2),\ldots,\mathbf{h}(T))
\end{align}
Let $\hat{\mathbf{e}}_g$ be the integrated vector obtained based on the $g$\,th annotation vector, and $\mathbf{E}=[\hat{\mathbf{e}}_1,\hat{\mathbf{e}}_2,\ldots,\hat{\mathbf{e}}_G]^{\text T}$. Let ${\mathbf{e}}$ be the window-level d-vector, ${\mathbf{e}}$ is obtained by vectorising $\mathbf{E}^{\text T}$, and 
${\mathbf{e}}=\text{Vector}(\mathbf{E}^{\text T})=\text{Concat}(\hat{\mathbf{e}}_1,\hat{\mathbf{e}}_1,\ldots,\hat{\mathbf{e}}_G)$ where $\text{Vector}(\cdot)$ and $\text{Concat}(\cdot)$ refer to the vectorisation and concatenation operations.
%correspondingly. 

To encapsulate more diverse information in $\hat{\mathbf{e}}$ by encouraging different output heads to generate more dissimilar annotation vectors, a penalty  term $\mu{\|\mathbf{A}^{\text T}\mathbf{A} - \mathbf{\Lambda}\|}^2_\text{F}$ can be minimised during training,  
%in Eq.~(\ref{eq:penalty}) is minimised during training.
%\begin{equation}
%P = \mu{||\mathbf{A}^T\mathbf{A} - \mathbf{\Lambda}||}^2_F 
%  = \mu\sum_i^h(\mathbf{a_i}^T\mathbf{a_i}-\lambda_i)^2 + \sum_{\substack{i,j, i\neq j}}^h(\mathbf{a_i}^T\mathbf{a_j})^2,
%\label{eq:penalty}
%\vspace{-0.3cm}
%\end{equation}
where $\|\cdot\|_{\text F}$ denotes the Frobenius norm, $\mathbf{\Lambda}$ is a diagonal coefficient matrix, and $\mu$ is the pre-defined penalisation coefficient. Each diagonal value of  $\mathbf{\Lambda}$, $\lambda_g=\mathbf{\Lambda}_{gg}$, controls the smoothness of the distribution of the values of the $g$\,th annotation vector. 
The penalty term encourages different values within each annotation vector to be orthogonal and their L2-norm values to approach the corresponding smoothness coefficient $\lambda$. In the original work on self-attentive structures \citep{selfattent0}, the values of all $\lambda$ are set to 1, which results in only sparse annotation vectors with spiky values. In our previous work \citep{mypaper}, a modified penalty term is proposed and showed that it is useful to also set some of the $\lambda$ values to be $1/G$ for extracting window-level d-vectors for diarisation, which encourages the values within the corresponding annotation vector to be more evenly distributed. 

\subsubsection{2D self-attentive combination}
\begin{figure}[h]
	\centering
		\includegraphics[scale=.4]{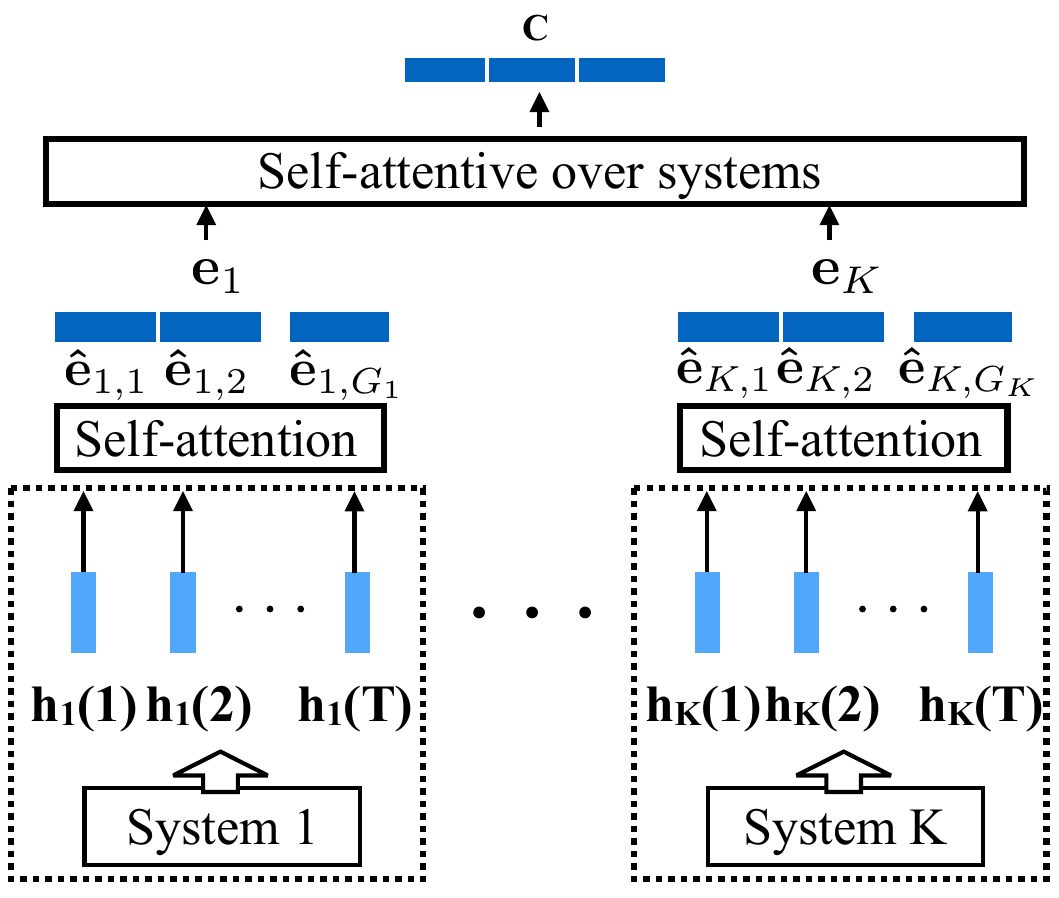}
	\caption{A sketch map of the 2D self-attentive combination applied to $K$ window-level d-vector extraction systems.}
	\label{FIG:2dselfatten}
\end{figure}

Next, the proposed 2D self-attentive combination is presented.
As illustrated in Fig.~\ref{FIG:2dselfatten}, in order to combine $K$ window-level d-vector extraction systems, a multi-head self-attentive structure is first used to generate the window-level d-vector for each system separately, and an additional multi-head self-attentive structure is then used to fuse different window-level d-vectors across systems.
%thereafter. 
Two different methods for 2D self-attentive combination are investigated, which differ in the input of the additional self-attentive structure for window-level d-vector combination.
\begin{itemize}
\item First, the same dynamic weight can be assigned to the integrated vectors derived from every output head of the same extraction system. That is,
%the self-attentive structure can be used to directly combine the window-level d-vectors as
%can be applied to the window-level d-vector $\hat{\mathbf{e}}$ combination could be performed on the multi-head output where all heads in the d-vector share the same annotation vector, as shown in Eq.~(\ref{eq:consec1}).
\begin{equation}
\mathbf{C}=\text{SelfAtt}\left(\mathbf{W}_1{\mathbf{e}}_1,\mathbf{W}_2{\mathbf{e}}_2\dots,\mathbf{W}_K{\mathbf{e}}_K\right),
%\mathbf{C}_{\text{consec1}}=\text{SelfAtten}\left(\hat{\mathbf{e}}_1,\hat{\mathbf{e}}_2\dots,\hat{\mathbf{e}}_k\right),
\label{eq:consec1}
\end{equation}
where ${\mathbf{e}}_k$ is the d-vector relevant to the $k$\,th extraction system. %The  $\mathbf{C}$ refers to the c-vector. 
$\mathbf{W}_k$ is a weight matrix used to permute the orders of components, in case that the d-vectors extracted by different pre-trained systems may not have the most suitably ordered vector components to be combined by the weighted sum of the attention mechanism. 
%In this method, the integrated vectors derived based on the different output heads of the same extraction system are scaled by the same dynamic weight. 

%\item Alternatively, it could be performed at the head level where different heads from the same system can be assigned different weights, 
\item Alternatively, a different dynamic weight can be assigned to every integrated vector. That is, 
%as shown in Eq.~(\ref{eq:consec2}). This relaxes the constraint on the number of heads for each system which has to be equal in the previous combination method. 
\begin{equation}
\mathbf{C}=\text{SelfAtt}\big(\mathbf{W}_1\hat{\mathbf{e}}_{1,1},\mathbf{W}_1\hat{\mathbf{e}}_{1,2}, \dots,\mathbf{W}_K\hat{\mathbf{e}}_{K,G_K}\big),
%\mathbf{C}=\text{SelfAtt}\big(\mathbf{W}_1\mathbf{e}_{1,1},\dots,\mathbf{W}_1\mathbf{e}_{1,h_1}, \dots,\mathbf{W}_k\mathbf{e}_{k,1},\dots,\mathbf{W}_k\mathbf{e}_{k,h_k}\big),
%\mathbf{C}_{\text{consec2}} = \text{SelfAtten}\big(\mathbf{e}_{11}, \dots,\mathbf{e}_{1h}, \dots,\mathbf{e}_{k1}, \dots,\mathbf{e}_{kh}\big),
\label{eq:consec2}
\end{equation}
where $\hat{\mathbf{e}}_{k,g}$  and $G_k$ are the integrated vector related to the $g$\,th output head and the number of output heads of the $k$\,th system respectively. It is clear that this method allows different systems to have different numbers of output heads. 
\end{itemize}
In both Eqns.~\eqref{eq:consec1} and \eqref{eq:consec2}, the c-vector, $\mathbf{c}$, is obtained as the vectorised version of $\mathbf{C}$.  

%, as speaker characteristics are usually encoded in different order by different systems, a fully-connected (FC) layer is introduced to transform each system output to a similar space before performing the self-attentive combination. 

Moreover, a baseline d-vector combination method is also introduced here. It first concatenates different d-vectors and then transforms the resulted vector with a ReLU FC layer
\begin{align}
\mathbf{c}&=\text{ReLU}\big(\mathbf{W}\,\text{Concat}({\mathbf{e}}_1,{\mathbf{e}}_2,\dots,{\mathbf{e}}_K)+\mathbf{b}\big),\label{eq:fcfusion}
    %\mathbf{c}&=\text{ReLU}\big(\mathbf{W}^{\text T}\text{Concate}({\mathbf{e}}_1,{\mathbf{e}}_2,\dots,{\mathbf{e}}_K)+\mathbf{b}\big),\label{eq:fcfusion}
\end{align}
where $\mathbf{W}$ and $\mathbf{b}$ are the weight matrix and the bias vector of the FC layer. 
%\begin{align}
%    \mathbf{C}_{\text{FCFusion}} & = f\big(\mathbf{W}^T[\mathbf{E}_1, \dots, \mathbf{E}_i, \dots, \mathbf{E}_k]\big),\label{eq:fcfusion}\\
%    \mathbf{C}_{\text{DirectSum}}& = f\big(\sum\nolimits_{i=1}^k\mathbf{W}_i^T\mathbf{E}_i\big),
%    \label{eq:directsum}
%\end{align}
%\begin{align}
%    \mathbf{C}_{\text{FCFusion}} & = f\big(\mathbf{W}^T[\mathbf{E}_1, \dots, \mathbf{E}_i, \dots, \mathbf{E}_k]\big),\label{eq:fcfusion}\\
%    \mathbf{C}_{\text{DirectSum}}& = f\big(\sum\nolimits_{i=1}^k\mathbf{W}_i^T\mathbf{E}_i\big),
%    \label{eq:directsum}
%\end{align}
%where $f$ is an arbitrary activation function and ReLU is used in experiments for easier clustering tuning.

\subsection{Gated additive combination}
\label{ssec:gating}
\begin{figure}[h]
    \centering
    \includegraphics[scale=0.45]{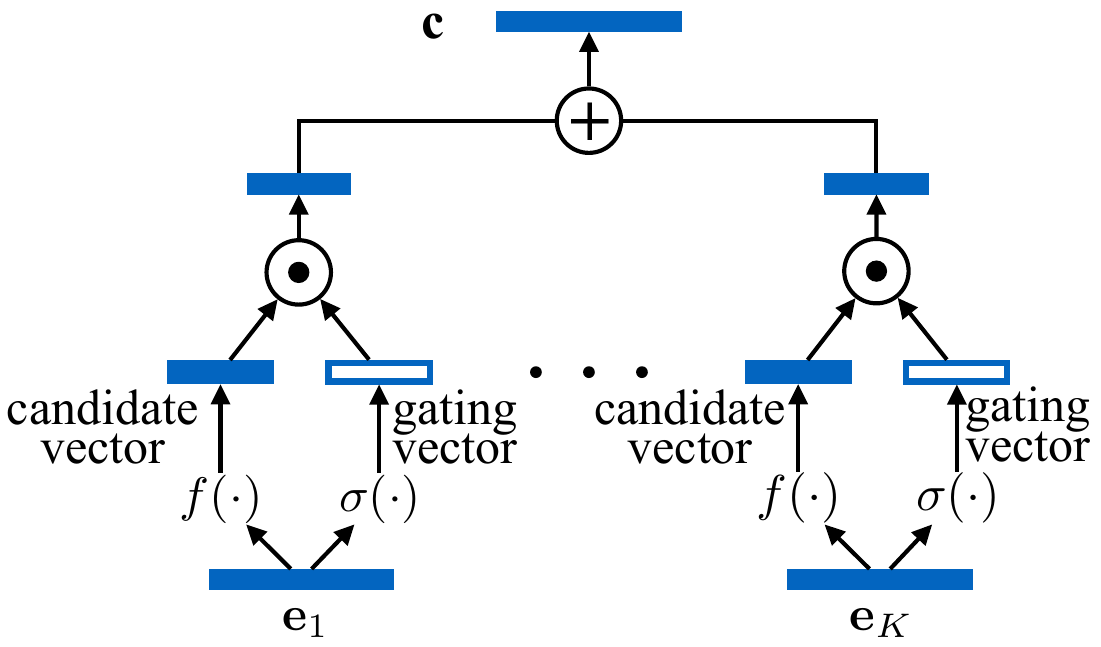}
    \caption{A sketch map of the gated additive combination for $K$ systems. $\sigma$, $\odot$, and $\oplus$ refer to the Sigmoid activation function, the Hadamard product of two vectors, and the element-wise vector addition respectively.}
    \label{fig:gating}
\end{figure}

The proposed gated additive combination structure relies on the gating mechanism as illustrated in Fig \ref{fig:gating}. Compared to the 2D self-attentive combination which scales all elements in each window-level d-vector by the same dynamic weight, gated additive combination allows each element in each candidate vector to be scaled with a different dynamic weight, before combination via vector addition.
%This combination is performed across speaker embeddings, $\mathbf{E}_i$, extracted using self-attentive layers across time from different systems. 
Analogous to an array of logic gates in electronics, in the gating mechanism \citep{lstm}, a gating vector is often derived as the output vector of an FC layer with Sigmoid activation functions, whose elements are used as the dynamic weights with soft 0-1 values. Similar to the 2D self-attentive combination in Eqns.~\eqref{eq:consec1} and \eqref{eq:consec2}, a separate FC layer is used to permute the order of the elements in each window-level d-vector to generate the candidate vectors. 

Specifically, a c-vector can be obtained by combining $K$ window-level d-vectors using gated additive combination as
\begin{equation}
    \mathbf{c}=\sum\nolimits_{k=1}^{K}f\big(\mathbf{W}_k\mathbf{e}_k+\mathbf{b}_{w,k}\big)\odot\sigma\big(\mathbf{U}_k\mathbf{e}_k+\mathbf{b}_{u,k}\big),
    \label{eq:gating}
\end{equation}
where $\mathbf{W}_k$, $\mathbf{U}_k$, $\mathbf{b}_{w,k}$, and $\mathbf{b}_{u,k}$ are the weight matrices and bias vectors of the FC layers used to derive the $k$\,th candidate vector and gating vector respectively.
$\sigma$ and $\odot$ denote the Sigmoid activation function and Hadamard product. $f(\cdot)$ is some activation function chosen for the candidate vector. 
%element-wise product and $f$ represents any activation function. Finally, the c-vector for classification and clustering, $\mathbf{C}_{gating}$, is obtained by summing up gated speaker embeddings.

\subsection{Bilinear pooling combination}
\label{ssec:bilinear}
Bilinear pooling \citep{bilinear_origin} is a commonly used approach for fusing multimodal representations \citep{multimodalreview}, which combines two vectors using the bilinear form 
%on their vector space by
\begin{equation}
    c_o=\mathbf{e}_1^{\text{T}}\mathbf{W}_o\,\mathbf{e}_2+b_o,
    \label{eq:bilinear_orig1}
\end{equation}
where $\mathbf{e}_1$ and $\mathbf{e}_2$ are the $M$-dimensional (-dim) and $N$-dim vectors to combine, and $\mathbf{W}_o$ and $b_o$ are the $M\times N$-dim weight matrix and the bias value corresponding to $c_o$. When generating an $O$-dim vector $\mathbf{c}=[c_1,c_2,\ldots,c_O]^{\text T}$ using bilinear pooling, the weight matrices $\mathbf{W}_1,\mathbf{W}_2,\ldots,\mathbf{W}_O$ form an $M\times N\times O$-dim weight tensor. It is equivalent to combining $\mathbf{e}_1$ and $\mathbf{e}_2$ using the vector outer product, and projecting the vectorisation of the resulting matrix into an $O$-dim vector space using a linear FC layer. That is,
\begin{equation}
    \mathbf{c}=\mathbf{W}\,\text{Vector}\big(\mathbf{e}_1\otimes\mathbf{e}_2\big)+\mathbf{b},
    \label{eq:bilinear_orig0}
\end{equation}
where $\otimes$ is the outer product, and $\mathbf{W}$ and $\mathbf{b}$ are the $(M\times N)\times O$-dim weight matrix and bias vector of the FC layer, with $\mathbf{W}=[\text{Vector}(\mathbf{W}_1),\text{Vector}(\mathbf{W}_2),\ldots,\text{Vector}(\mathbf{W}_O)]^{\text T}$.
Therefore, in contrast to the attention-based and gating-based methods in Sections~\ref{ssec:selfatt} and \ref{ssec:gating} that combine vectors in the $M$-dim or $N$-dim representation space, bilinear pooling combines two vectors by computing their outer product to capture the multiplicative interactions between all possible element pairs in a more expressive $M\times N$-dim space, which is projected to another $O$-dim vector space with $\mathbf{W}$. %Furthermore, although the original bilinear pooling is only applicable for combining two vectors, it is straightforward to extend Eqn.~\eqref{eq:bilinear_orig0} to the case with $K$ vectors as
%\begin{equation}
%    \mathbf{c}=\mathbf{W}\,\text{Vector}\bigger(\mathbf{e}_1\otimes\mathbf{e}_2\ldots\otimes\mathb f(\mathbf{e}_K)\bigger)+\mathbf{b},
%    \label{eq:bilinear_orig2}
%\end{equation}

%Thus the parameters required by bilinear pooling to generate the output vector $\mathbf{c}$ include a weight tensor and a bias vector whose dimensions are $M\times N\times O$ and $O$ respectively. 

Although bilinear pooling is a powerful vector combination method in theory, in practice it often suffers from issues caused by its high dimensionality (typically
on the order of hundreds of thousands to a few million
dimensions) that
requires decomposition of the weight tensor to allow the associated parameters to be estimated properly and efficiently. Commonly used bilinear pooling decomposition methods include count sketches and convolutions \citep{MCB1}, low-rank approximations \citep{bilinear}, and various other tensor decomposition methods \citep{MCB2,MUTAN,BLOCK}. In this section, a modified multimodal low-rank bilinear attention network \citep{bilinear} is proposed to combine two window-level d-vectors.

\citet{LowRankBilinear} suggested the use of a low-rank approximation $\mathbf{W}_o\approx\mathbf{U}_{o1}\mathbf{U}_{o2}^{\text T}$ for bilinear pooling, where $\mathbf{U}_{o1}$ and $\mathbf{U}_{o2}$ are ${M\times D}$-dim and ${N\times D}$-dim matrices, and $D$ is the reduced rank of $\mathbf{W}_o$ which is smaller than or equal to the minimum value between $M$ and $N$. Eqn.~\eqref{eq:bilinear_orig1} can be rewritten as $c_o\approx\mathbf{e}_1^{\text{T}}\mathbf{U}_{o1}\mathbf{U}_{o2}^{\text T}\,\mathbf{e}_2+b_o=\mathbf{1}^{\text T}(\mathbf{U}_{o1}^{\text T}\mathbf{e}_1\odot\mathbf{U}_{o2}^{\text T}\mathbf{e}_2)+b_o$, where $\odot$ still refers to the Hadamard product and $\mathbf{1}$ is a $D$-dim vector whose elements are all set to one. 
Although the low-rank approximation can reduce the number of parameters, it still relies on the use of an $M\times D\times O$-dim tensor and an $N\times D\times O$-dim tensor formed by $\mathbf{U}_{11},\mathbf{U}_{21},\ldots,\mathbf{U}_{O1}$ and $\mathbf{U}_{12},\mathbf{U}_{22},\ldots,\mathbf{U}_{O2}$ respectively. To remove the use of tensors, \citet{bilinear} proposed to tie all $\mathbf{U}_{o1}$ matrices as $\mathbf{U}_1$ and all $\mathbf{U}_{o2}$ matrices as $\mathbf{U}_2$, and a $D$-dim vector $\mathbf{p}_o$ is used to distinguish the value of $c_o$ from the other elements of $\mathbf{c}$ by
$c_o\approx\mathbf{p}_{o}^{\text T}(\mathbf{U}_1^{\text T}\mathbf{e}_1\odot\mathbf{U}_2^{\text T}\mathbf{e}_2)+b_o$.
It has been also shown that using non-linear activation functions to transform the vectors before the Hadamard product
can often result in better-performing models \citep{bilinear}. Hence the initial combination of d-vectors obtained using low-rank bilinear pooling with the Hadamard product is
\begin{equation}
    \mathbf{c}^{*}=\mathbf{P}\left(f(\mathbf{U}_1^{\text{T}}\mathbf{e}_1)\odot f(\mathbf{U}_2^{\text{T}}\mathbf{e}_2)\right)+\mathbf{b},
    \label{eq:bilinear}
\end{equation}
where $\mathbf{P}=[\mathbf{p}_1,\mathbf{p}_2,\ldots,\mathbf{p}_O]^{\text T}$ is an $O\times D$-dim projection matrix, $f(\cdot)$ is an activation function with a bounded range of output values, such as $\sigma(\cdot)$ and $\text{tanh}(\cdot)$. Furthermore, although a general modification of Eqn.~\eqref{eq:bilinear} with shortcut connections was given, shortcut connections were actually not used in \citep{bilinear}. In this paper, shortcut connections are included in bilinear pooling by using
\begin{equation}
    \mathbf{c}=\mathbf{c}^{*}+ \mathbf{V}_1\mathbf{e}_1+ \mathbf{V}_2\mathbf{e}_2,
    \label{eq:bilinear_short}
\end{equation}
where $\mathbf{V}_1$ and $\mathbf{V}_2$ are $O\times M$-dim and $O\times N$-dim projection matrices used to create the shortcut connections.
We found in our experiments that the shortcut version of low-rank bilinear pooling given in Eqn.~\eqref{eq:bilinear_short}
 outperformed the widely used form given in Eqn.\eqref{eq:bilinear}.

\begin{figure}[h]
    \centering
    \includegraphics[scale=0.5]{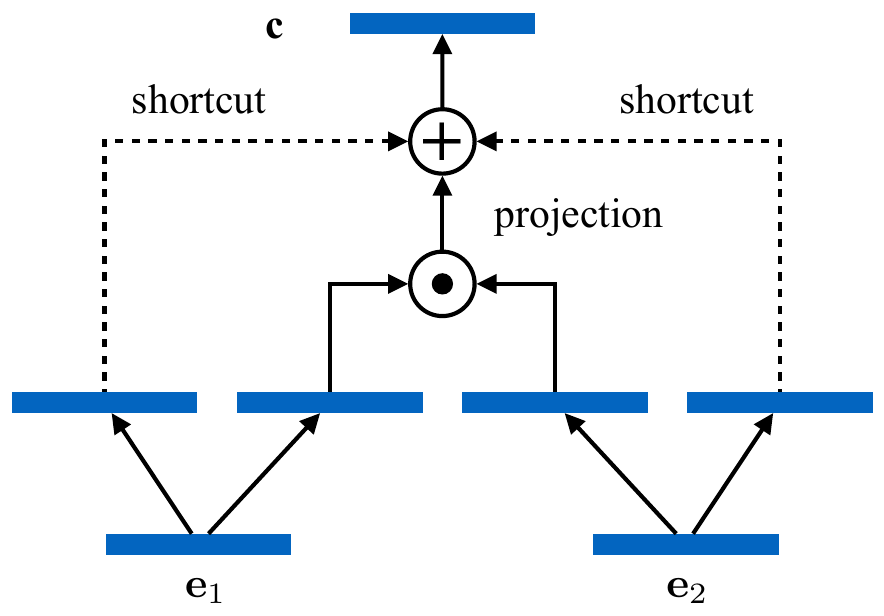}
    \caption{A sketch map of the bilinear pooling combination over two d-vector systems. $\odot$ represents the Hadamard product and $\oplus$ represents the element-wise addition of vectors.}
    \label{fig:bilinear}
\end{figure}

Moreover, although bilinear pooling is only applicable for combining two vectors, a similar idea using vector outer product to combine $K$ $(K>2)$ d-vectors can be achieved by extending 
Eqn.~\eqref{eq:bilinear_orig0} as
\begin{equation}
    \mathbf{c}=\mathbf{W}\,\text{Vector}\big(\mathbf{e}_1\otimes\mathbf{e}_2\ldots\otimes\mathbf(\mathbf{e}_K)\big)+\mathbf{b},
    \label{eq:bilinear_orig2}
\end{equation}
and Eqns.~\eqref{eq:bilinear} and \eqref{eq:bilinear_short} can be extended similarly as
\begin{align*}
\mathbf{c}^{*}&=\mathbf{P}\left(f(\mathbf{U}_1^{\text{T}}\mathbf{e}_1)\odot f(\mathbf{U}_2^{\text{T}}\mathbf{e}_2)\cdots\odot f(\mathbf{U}_K^{\text{T}}\mathbf{e}_K)\right)+\mathbf{b}\\
%\label{eq:bilinear2}
\mathbf{c}&=\mathbf{c}^{*}+\sum\nolimits_{k=1}^{K}\mathbf{V}_k\mathbf{e}_k.
%\label{eq:bilinear_short2}
\end{align*}

%Last but not least, previously mentioned c-vectors obtained via additive methods can be further combined with other d-vectors using the bilinear pooling combination to exploit the complementary between d-vectors and c-vectors and between different combination methods. In particular, the 2D self-attentive c-vector is used in this paper together with a subset of d-vectors it combines to perform the bilinear pooling combination as an example, as shown in Fig.
\subsection{Stacked combination}
\label{ssec:stacked}
Lastly, a stacked combination method is proposed to exploit the use of the complementarity among different vector combination structures. There are multiple possible designs for such structures, and we only focus on those that stack any two among the three proposed combination methods.
%focus on those stacking any two among the 2D self-attentive combination, the gated additive combination, and the bilinear pooling combination.

%Lastly, the three c-vector generation methods proposed in this section can be combined to further exploit the complementarity among different vector combination structures. 
%There are many possible ways to design such structures, and in this paper, 

\begin{figure}[h]
    \centering
    \includegraphics[scale=0.5]{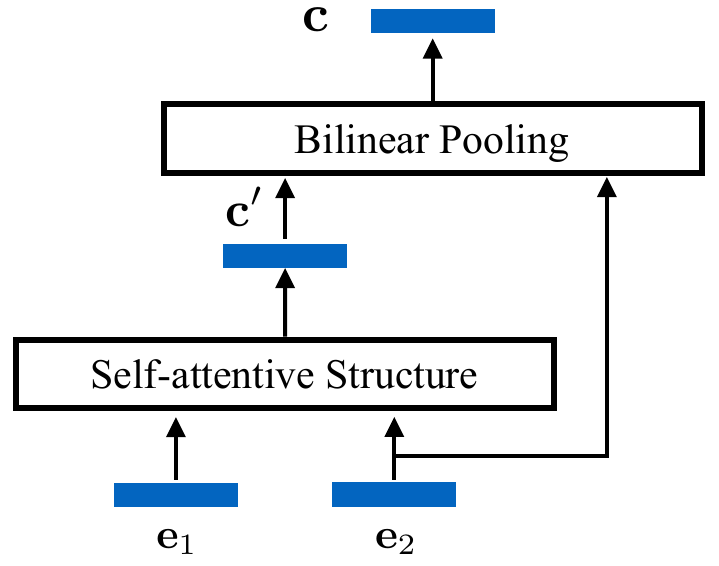}
    \caption{A sketch of the stacked combination method consists of the first type of 2D self-attentive combination and bilinear pooling combination over two d-vectors. $\mathbf{c}'$ is the initial c-vector obtained using 2D self-attentive combination.}
    \label{fig:doublecomb}
    \vspace{-0.5cm}
\end{figure}

Here we propose a structure stacking the 2D self-attentive combination and the bilinear pooling combination, which is found to perform the best in our experiments. Specifically, the two d-vectors, $\mathbf{e}_1$ and $\mathbf{e}_2$, are first combined through the first type of 2D self-attentive combination to generate the initial c-vector $\mathbf{c}'$, and then $\mathbf{c}'$ is fused with $\mathbf{e}_2$ again using the bilinear pooling to generate the final c-vector, $\mathbf{c}$. This stack of combinations is illustrated in Fig.~\ref{fig:doublecomb}.
%previously mentioned c-vectors obtained via additive methods can be further combined with other d-vectors using the bilinear pooling combination to exploit the complementary between d-vectors and c-vectors and between different combination methods. 
%In particular, the 2D self-attentive c-vector is used in this paper together with a subset of d-vectors it combines to perform the bilinear pooling combination as an example, as shown in Fig.\ref{fig:doublecomb}. 
\section{Experimental setup}
\label{sec:expsetup}
\subsection{Dataset details}
All of the data preparation and model training was done using an extended version of HTK version 3.5.1 and PyHTK \citep{htk, pyhtk}. All systems were trained on the augmented multi-party interaction (AMI) meeting corpus. 
The full AMI training set contains 135 meetings with 155 speakers recorded, of which, 10\% of the data for each speaker was used for held-out validation during training. The development (\textbf{Dev}) and evaluation (\textbf{Eval}) sets from the AMI official speech recognition partition were used to evaluate the performance of the proposed methods, whose details are shown in Table \ref{tab:data}. The training set has 4--5 speakers per meeting while Dev and Eval have 4 speakers in every meeting. %Comparing to other studies that often uses a selection of meetings for diarisation test \citep{**}, our work directly uses the original speech recognit 
\begin{table}[h]
    \centering
    \caption{Details of the AMI official speech recognition partition.}
    \begin{tabular}{ccc}
    \toprule
        Dataset &  \#Meetings & \#Speakers\\
         \midrule
        Train & 135 & 155 \\
        Dev & 18 & 21 (2 in Train)\\
        Eval & 16 & 16 (0 in Train)\\
        \bottomrule
    \end{tabular}
    \label{tab:data}
\end{table}

The input features to all systems are from a 40-dim log-Mel filter bank with 25 millisecond frame size and 10 millisecond frame increment. The acoustic features were extracted from the multiple distance microphone (MDM) audio data processed by beam-forming using the BeamformIt toolkit \citep{rt05paper,beamform}. Phase information is not included in the feature as the focus of this paper is the modelling technique \citep{beamform,Phase2,phase}. 
Furthermore, the NIST rich transcription evaluation 2005 dataset (\textbf{RT05}) is used as an additional testing set, which consists of meetings with 4--10 speakers from various meeting corpora. 

It is worth noting that although AMI is widely used for speaker diarisation, most studies either use different sets of meetings for testing, or the independent headset microphone audio, such as \citep{e2eoverlap,setup1,setup2}. This makes the results not comparable to those presented in this paper. %The work in \citep{AMIlarge,speakerGAN} used the official speech recognition data partition and MDM audios on manual segmentation assuming ideal VAD. 
\citet{AMIlarge,speakerGAN} used the official speech recognition data partition and MDM audio with manual segmentation. %, which makes the results SERs. 
%While the SERs presented in these papers do not reflect real performance difference of the methods since different training sets were used, the best SER of 7.9\% produced by 
%the best-performing system achieved a similar SER to their best-reported numbers. 
However, the SERs in their papers are not comparable to the DERs in Table~\ref{tab:voxceleb} since the systems were trained with different training data and tested differently. 
\citet{AMIlarge3} used automatic segmentation for their experiments on AMI, but the scoring procedure is different from ours since it does not include any collar.

%More recent studies, such as \citep{e2eoverlap}, use Dev and Eval as the diarisation test sets, which include more diversified meetings collected at different universities. Compared to \citep{e2eoverlap}, this paper uses the MDM data recorded by real far-field microphone arrays instead of the artificially created mixture of audios from independent near-field headset microphones. To the best of the authors' knowledge, %although there is no study using the same setup for AMI speaker diarisation as ours to date,
%although the setup and results used in other AMI speaker diarisation studies are not comparable to ours,
%our system can still have superior performance, since the results in Table~\ref{tab:finalresult} are the lowest error rates with the same training data, provided that our more realistic setup can increase the difficulty and result in higher error rates.

\subsection{System specifications}
To extract window level d-vectors, a 2-second sliding window was applied with a 1-second overlap between adjacent windows. The TDNN and the HORNN with ReLU activation functions were chosen as two example systems for d-vector extraction and combination. The details of the TDNN structure is shown in Table~\ref{tab:tdnnspec}, which resembles the one used in \citep{xvec}. 
%TDNN system resembles the structure proposed in the x-vector system, except the statistical pooling layer is replaced by the self-attentive layer. 
The HORNN system has two recurrent connections to both the previous hidden state and that 4 time steps before the current time step, which provides a simple solution to the gradient vanishing issue by using fewer parameters and calculations than LSTM. The HORNN model has 256-dim hidden states that are projected to 128-dim vectors using a shared linear projection \citep{hornn}. 
%All activation functions for the hidden layers of each system are ReLUs. 
%The TDNN structure is shown in Table. \ref{tab:tdnnspec}, and for a similar performance, the HORNN has 256d hidden states 128d projection dimension. 
%Next, for either system, the 128-dim output vectors from TDNN or HORNN are mapped to 128-dim before being fed into the corresponding self-attentive structure with 5 heads for temporal pooling. 
%Next, for either system, a 
A 5 head self-attentive structure was then used for temporal pooling across the 128-dim frame-level d-vectors extracted by either a TDNN or a HORNN model. As discussed in Section~\ref{sssec:selfattentive}, the $\lambda$ values for the penalty term related to each head are set to 1, 1, 1, 0.2, and 0.2. 
Furthermore, instead of including every frame in the 2 second window as for the TDNN model, the self-attentive structure in the HORNN model samples one frame-level d-vector for every 10 frames that considerably improves the computational efficiency without significant performance degradation. In other words, the lengths of the input sequences to the self-attentive structures of TDNN and HORNN are 200 and 20 respectively. The $640$-dim ($128\times5$) window-level d-vectors from the TDNN and HORNN are denoted $\mathbf{e}_{\text T}$ and $\mathbf{e}_\text{H}$ and used in combination. The notation for different combination methods used in the rest of the paper are listed in Table~\ref{tab:notation}.
%The combination methods given in Eqns.~\eqref{eq:consec1}, \eqref{eq:consec2}, \eqref{eq:fcfusion}, and \eqref{eq:gating} are in correspondence with the systems named ``2D Self-atten. 1'', ``2D Self-atten. 2'', ``FC Fusion'', and ``Gated Add.'' respectively later in Section~\ref{sec:results}. ``Bilinear (Sigmoid)'' and ``Bilinear (Tanh)'' refer to bilinear pooling  and $f(\cdot)=\sigma(\cdot)$ and $f(\cdot)=\text{tanh}(\cdot)$ separately in Eqn.~\eqref{eq:bilinear_short}.
%Then, two 640-dim system output vectors are sent to the combination stage to extract c-vectors. Details of the combination stage can be found in Append. \ref{append1}.

\begin{table}[h]
    \centering
        \caption{The structure of a TDNN model used for extracting one frame-level d-vector at time step t. The outputs from the required contexts of a layer are concatenated as the inputs to the next layer. 
        %The first five layers operate at frame-level, with a small temporal context centred at the current frame t. 
        ``Layer'' includes the index and the activation function of the layer, ``\#Frames'' shows the total number of frames being used as the input when reaching that layer, and ``Dimensions''
        gives the input dim.$\times$ output dim. of each layer.}
    \begin{tabular}{lccc}
    \toprule
       Layer &  Context & \#Frames & Dimensions \\
       \midrule
       1, ReLU&  t-2,t-1,t,t+1,t+2 & 5 & 200 x 256  \\
        2, ReLU & t-2,t,t+2 & 9 & 768 x 256\\
        3, ReLU & t-3,t,t+3 & 15 & 768 x 256\\
        4, ReLU &  t & 15 & 256 x 128\\
        \bottomrule
    \end{tabular}
    \label{tab:tdnnspec}
\end{table}

\begin{table}[h]
    \centering
    \caption{Notation and description of different combination methods for d-vectors $\mathbf{e}_{\text {T}}$ and $\mathbf{e}_{\text {H}}$.}
    \begin{tabular}{ll}
    \toprule
        System & Description \\
         \midrule
         FCFusion$(\mathbf{e}_{\text {T}},\mathbf{e}_{\text {H}})$ & FC layer combination using Eqn.~\eqref{eq:fcfusion}\\
        SelfAtt1$(\mathbf{e}_{\text {T}},\mathbf{e}_{\text {H}})$ & The first type of 2D self-attentive \\
        & combination using Eqn.~\eqref{eq:consec1} \\
        SelfAtt2$(\mathbf{e}_{\text {T}},\mathbf{e}_{\text {H}})$ & The second type of 2D self-attentive \\
        & combination using Eqn.~\eqref{eq:consec2} \\
        GatedAdd$(\mathbf{e}_{\text {T}},\mathbf{e}_{\text {H}})$ & Gated add. combination with Eqn.~\eqref{eq:gating}\\
        Bilinear$_{f}(\mathbf{e}_{\text {T}},\mathbf{e}_{\text {H}})$ & Bilinear pooling 
        combination based on \\ & Eqn.~\eqref{eq:bilinear_short} with activation function $f(\cdot)$\\ 
        \bottomrule
    \end{tabular}
    \label{tab:notation}
\end{table}

After the combination stage, another linear FC layer  was used to project the derived c-vectors, $\mathbf{c}$, down to a 128-dim space, which was then used as the final input to the clustering algorithm. 
Both the TDNN and HORNN models were pre-trained in a layer-by-layer fashion without self-attentive structures. 
%window-level d-vector systems are initialised by  frame-level pre-training, and then jointly trained in the combination networks. 
Finetuning was performed by jointly training all parameters of each window-level d-vector extraction system.
When performing c-vector combination, unless explicitly stated. the combination parameters were also jointly trained with the parameters associated with the window-level d-vector extraction systems. 
Moreover, instead of using the standard softmax output activation function with cross-entropy loss at the output layers of the d-vector and c-vector extraction systems, the angular softmax training loss \citep{softmax4,asoftmax} was adopted with the $m$ factor set to $1$, 
to ensure that the derived embeddings are trained to discriminate speakers based on the cosine distance.  
%better angular  discrimination. 
%This further helps the clustering process as spectral clustering is based on cosine distances.
This improves consistency between the performance of the speaker embeddings and the clustering results, since we use spectral clustering with the cosine distance. 

\subsection{Evaluation metrics}
Model performance was evaluated using the diarisation error rate (DER) which is the sum of the speaker (clustering) error rate (SER), missed speech (MS) and false alarm (FA). As in training, a 2-second sliding window with 1-second overlap is applied, and 128-dim c-vectors were extracted as the output vectors from the penultimate layer. 
%by forward propagation to the bottleneck layer. 
%The threshold value for spectral clustering pre-processing stage is tuned on dev set, and directly applied to eval set. Scoring uses the setup from the NIST rich transcription evaluations with a 0.25 second collar. Furthermore, by comparing the oracle segmentation with the alignment file, non-speech intervals longer than 0.4 seconds which were original included were removed from the reference file while leaving 0.2 second collar on both ends to tolerate alignment errors. This results in a new reference file with fewer non-speech parts that is more suitable for diarisation evaluation. Evaluations against both original and modified reference files are presented.
The threshold value for spectral clustering pre-processing stage is tuned on the Dev set, and directly applied to the AMI Eval set and RT05. Scoring uses the setup from the NIST rich transcription evaluations with a 0.25 second collar. If not specified, overlapped speech was not scored.
\begin{table}[h]
    \caption{Statistics of the original and modified reference files. ``\#Segments'' shows the number of scored speech segments in the reference. ``\%Overlap'' is measured by the total time of overlap speech divided by the total segmented audio time. ``Time'' gives the actual time after applying collars and removing overlap, and the total intra-segment non-speech time.}
    \begin{tabular}{lccc}
    \toprule
       Reference  & \#Segments & \%Overlap & Time (second)  \\
    \midrule
    Dev (Original) & 13059 & 15.2 & 18613 / 7290\\
     Dev (Modified)  & 17218 & 10.5 & 14973 / 3082 \\ 
     Eval (Original) & 12612 & 15.3 & 18075 / 8241\\
     Eval (Modified) & 17100& 9.6 & 14021 / 3134\\
     \bottomrule
    \end{tabular}
    \label{tab:references}
\end{table}
Apart from the original manual segmentation with the official AMI release (termed as the \textbf{original segmentation}), we created a modified version of the manual segmentation (termed as the \textbf{modified segmentation}) by comparing each original manual segment with frame to speech and non-speech alignments generated by forced alignment  using a pre-trained speech recognition system \citep{htk}. 
To form the modified segmentation, non-speech intervals that were longer than 0.2 seconds in the original segmentation were reduced to leave a 0.1 second collar on both ends of each segment. 
The manual segmentation was modified since many original manual segments had very long non-speech parts at their beginning or end, which results in many unnecessary overlapping regions and has an impact on the speaker clustering performance. The \textbf{original reference} was also modified accordingly to form the \textbf{modified reference} to match the modified segmentation.\footnote{original and modified references are available here \url{https://github.com/BriansIDP/AMI_diar_references.git}} 

Statistics of both the original and modified references, including the number of segments, the portion of overlapped speech, the total scored time, and the total intra-segment non-speech time are shown in Table~\ref{tab:references}. %In contrast to the original reference, 
In the modified reference for the Dev and Eval sets, both the total scored time and the total intra-segment non-speech time have been reduced. Since the reduction in the total intra-segment non-speech time is larger, the amount of speech scored in the modified reference is increased, which shows the benefit of modifying the reference.
Meanwhile, the total time of overlapped segments of Dev and Eval is reduced from 4,178 and 4,049 seconds to  2,535 and 2,180 seconds respectively.
\section{Experimental results}
\label{sec:results}
\subsection{AMI Results with Manual Segmentation}
This section gives the results of
%demonstrates the 
experiments performed on the AMI data  using the manual segmentation without using the VAD and CPD stages.
It directly compares the speaker clustering performance with different model structures for speaker embedding extraction.

\begin{table*}[width=2.0\linewidth, pos=t]
    % \centering
    \caption{AMI results in \%SER of different model combination methods on the manual segmentation. ``Original'' and ``Modified'' refer to the results obtained by testing on the original and modified segmentation and scored against the  original and modified reference files accordingly.
    ``\#Params.'' shows the number of parameters in each combination structure, and ``Classification Accuracy'' is the speaker classification accuracy on the held-out validation set. ``Manual selection per meeting'' is an oracle system obtained by manually switching between the TDNN and HORNN systems according to their \%SERs on Dev or Eval separately.} 
    %2D Self-atten 1 and 2 refers to c-vectors in Eqns. \eqref{eq:consec1} and \ref{eqeq:consec2} respectively. 
    %Activation functions before Hadamard product for bi-linear pooling are indicated in brackets.}
    \begin{tabular*}{\tblwidth}{@{} Lcccccc@{} }
    \toprule
    \multirow{2}{*}{System} & \#Params & Classification & \multicolumn{2}{c}{Original} & \multicolumn{2}{c}{Modified} \\
    \cmidrule(lr){4-5} \cmidrule(lr){6-7}
     & (million) & accuracy (\%) & Dev & Eval & Dev & Eval \\
    \midrule
        $\mathbf{e}_{\text {T}}$ (TDNN) & 0.6 & 77.6 & 14.3 & 15.4 & 19.8& 15.4 \\
        $\mathbf{e}_{\text {H}}$ (HORNN) & 0.2 & 74.1 & 13.2 & 15.4 & 13.6& 15.9 \\
        \midrule
        FCFusion$(\mathbf{e}_{\text {T}},\mathbf{e}_{\text {H}})$ & 1.0 & 76.4 & 12.8 &	14.7  & 13.5 & 15.2 \\
        SelfAtt1$(\mathbf{e}_{\text {T}},\mathbf{e}_{\text {H}})$ & 0.8 & 79.1 &  12.1	& 14.1 & 11.5 & 14.5 \\
        SelfAtt2$(\mathbf{e}_{\text {T}},\mathbf{e}_{\text {H}})$ & 0.8 & 78.6 & 12.0	& 13.1 & 11.8 & 13.9 \\
        GatedAdd$(\mathbf{e}_{\text {T}},\mathbf{e}_{\text {H}})$ & 1.0 & 80.0 & 12.8	& 12.3 & 12.1 & 11.7 \\
        Bilinear$_{\sigma}(\mathbf{e}_{\text {T}},\mathbf{e}_{\text {H}})$  & 1.0 & 80.7 & 11.6	& 12.9 & 11.9 & 14.2 \\
        Bilinear$_{\text{tanh}}(\mathbf{e}_{\text {T}},\mathbf{e}_{\text {H}})$ &  1.0 & 80.6 & 12.2 & 	13.6 & 12.9 & 13.9 \\
        \midrule
        Bilinear$_{\sigma}(\text{SelfAtt1}(\mathbf{e}_{\text {T}},\mathbf{e}_{\text {H}}),\mathbf{e}_{\text {H}})$ & 2.2 & 80.6 & \textbf{10.6} &	{12.0} & \textbf{10.7} & \textbf{10.5} \\
        Bilinear$_{\text{tanh}}(\text{SelfAtt1}(\mathbf{e}_{\text {T}},\mathbf{e}_{\text {H}}),\mathbf{e}_{\text {H}})$ & 2.2  & \textbf{80.9} &  11.4 &	12.5 &11.4 & 12.5\\
        \midrule
        Manual selection per meeting & N/A & N/A & 12.7 & \textbf{11.4} & 13.4 & 11.5\\
    \bottomrule
    \end{tabular*}
    \label{tab:manual}
\end{table*}

\subsubsection{Analysis on the penalty term of the multi-head self-attentive structure}
\begin{figure}[h]
    \centering
    \includegraphics[scale=0.3]{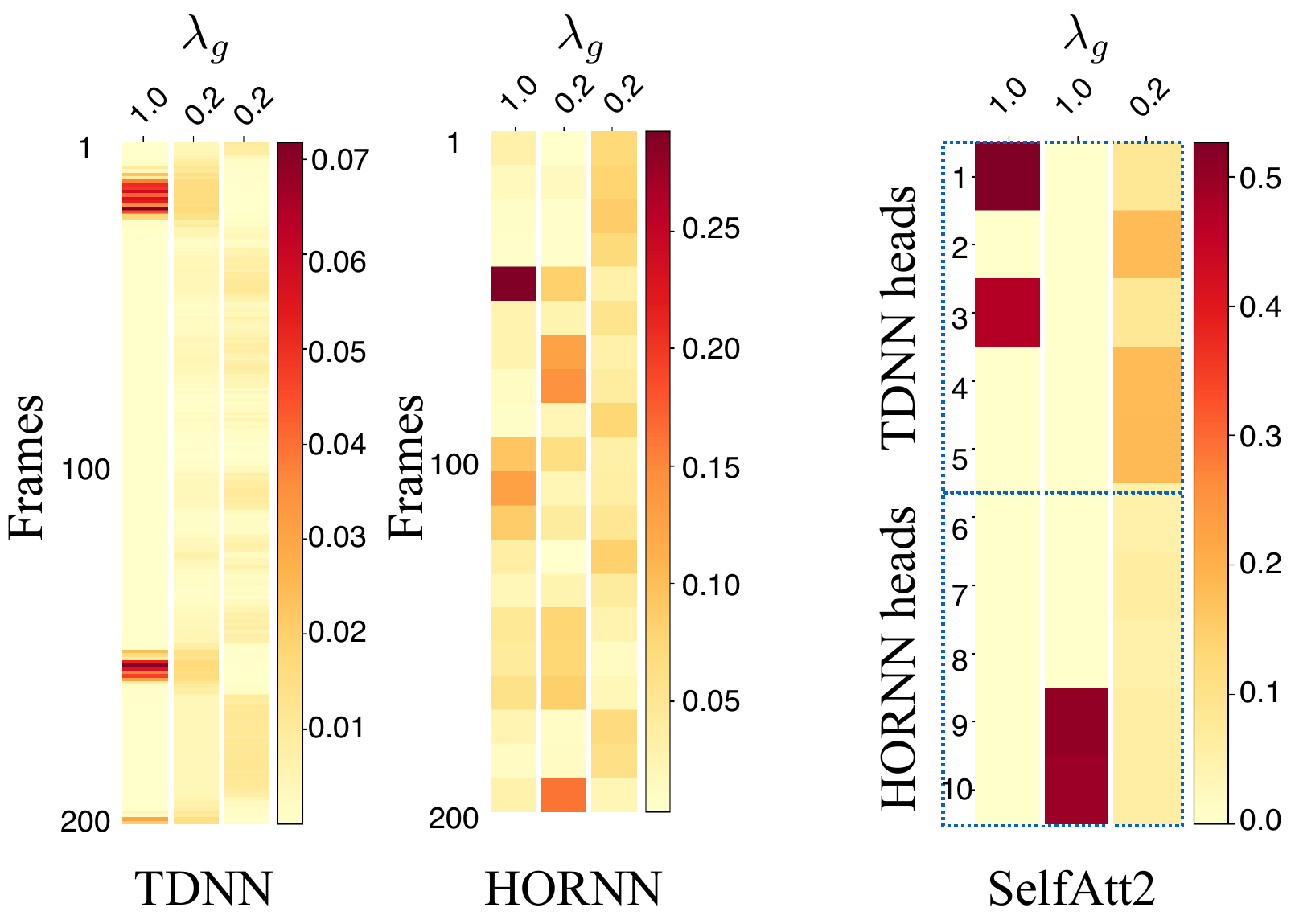}
    \caption{Heatmaps of the dynamic weights in selected annotation vectors produced by different self-attentive structures with the same input window. Left: Annotation vector values over frames that integrate frame-level d-vectors by both the TDNN and HORNN systems. Right: Annotation vector values assigned to each integrated vector $\hat{\mathbf{e}}_{k,g}$ by the second type of 2D self-attentive combination to generate a c-vector. Three annotation vectors relevant to different $\lambda_g$ values in the penalty term are selected from the system and shown in each plot.}
    \label{fig:attenhornn}
    \vspace{-0.5cm}
\end{figure}

%In addition to the combination weight distribution across time for TDNN system shown in \citep{mypaper}, a similar analysis on HORNN system is illustrated as shown in Fig \ref{fig:attenhornn}.
As discussed in Section~\ref{sssec:selfattentive} and shown in Fig.~\ref{fig:attenhornn}, $\lambda_g$, the self-attentive structure penalty term coefficient relevant to the $g$\,th annotation vector, can determine if the dynamic weights in the annotation vector has a ``spiky'' or ``smooth'' distribution. 
As found in our previous work \citep{mypaper}, mixing the ``spiky'' and ``smooth'' annotation vectors can result in better-performing speaker embeddings, and therefore is used throughout this paper. 

%The same smoothness control using $\lambda_g$ is also found in the dimension across integrated vectors from both systems. 
Furthermore, the plot in the right part of Fig. \ref{fig:attenhornn} shows that the annotation vectors focus on different integrated vectors from different systems, indicating the complementarity between the example TDNN and HORNN d-vector systems.

% The system of the first type of consecutive 2D self-attention, Self-atten. 1, uses a single annotation vector for window-level d-vector combination, whose dynamic weights are shown in Fig.\ref{fig:attenconsec1}.
% % \begin{figure}[h]
% %     \centering
% %     \includegraphics[scale=0.37]{Attenplot2.pdf}
% %     \caption{$C_{consec1}$ model combination weights, taking all 7 windows, $\text{W}_1$ to $\text{W}_7$, of a selected utterance.}
% %     \label{fig:attenconsec1}
% % \end{figure}
% The weight distribution for system combination exhibits a consistent fluctuations around a ratio close to 1, which reflects that the combination mechanism attaches similar but dynamic importance to each model. Besides, the combination slightly bias towards TDNN system as a result of its higher speaker classification accuracy.

\subsubsection{SER Comparison}
Next, the speaker clustering results using manual segmentation obtained using different combination methods are presented and discussed. 
%are shown in Table \ref{tab:manual}. 
As shown in Table \ref{tab:manual}, since the clustering threshold used for spectral clustering was determined based on Dev set performance, the Dev results are more consistent with the speaker classification accuracy on the validation set than those on Eval.  
The poorer TDNN performance on Dev set using the modified segmentation is due to two meetings having significantly higher SER than when using original segmentation.
%Since the threshold for clustering is determined based on the Dev set performance, results on the dev set are more consistent than that for the eval set. 
By comparing the results obtained with both the original and modified segmentation, apart from FCFusion$(\mathbf{e}_{\text {T}},\mathbf{e}_{\text {H}})$, all the other c-vector systems outperform both TDNN and HORNN d-vector systems. %This indicates that the baseline fusion method using an FC layer may not be powerful enough to capture the 
For the other c-vector systems, Bilinear$_{\sigma}(\mathbf{e}_{\text {T}},\mathbf{e}_{\text {H}})$ gives the lowest SERs with the original segmentation but not such good results for the modified segmentation. GatedAdd$(\mathbf{e}_{\text {T}},\mathbf{e}_{\text {H}})$ produces consistently low SERs with both segmentations. 

%According to 
From the results in Table \ref{tab:manual}, it is clear that the systems with multiplicative combination, Bilinear$_{\sigma}(\mathbf{e}_{\text {T}},\mathbf{e}_{\text {H}})$  and Bilinear$_{\text{tanh}}(\mathbf{e}_{\text {T}},\mathbf{e}_{\text {H}})$, generate different and  complementary outputs compared with the additive combination methods. 
%Therefore, an extra type of combination method that first combines the d-vectors from TDNN and HORNN usibng the first type of 2D self-attentive combination based on Eqn.~\eqref{eq:consec1}, and then combines the resulted c-vector with the  d-vector derived from HORNN again using bilienar pooling. 
The systems that combine the additive and multiplicative combination structures, Bilinear$_{\sigma}(\text{SelfAtt1}(\mathbf{e}_{\text {T}},\mathbf{e}_{\text {H}}),\mathbf{e}_{\text {H}})$ and
Bilinear$_{\text{tanh}}(\text{SelfAtt1}(\mathbf{e}_{\text {T}},\mathbf{e}_{\text {H}}),\mathbf{e}_{\text {H}})$, whose structures are shown in Fig.~\ref{fig:doublecomb}, use $\sigma(\cdot)$ and $\text{tanh}(\cdot)$ as the activation function $f(\cdot)$ respectively.  Bilinear$_{\sigma}(\text{SelfAtt1}(\mathbf{e}_{\text {T}},\mathbf{e}_{\text {H}}),\mathbf{e}_{\text {H}})$ 
 %Such a combination of the additive and multiplicative combination methods results in the systems of ``2D Self-atten. 1$+$Bilinear (Sigmoid)'' and ``2D Self-atten. 1$+$Bilinear (Tanh)''. The system of 2D Self-atten. 1$+$Bilinear (Sigmoid)
gives the lowest SERs with both types of segmentation and the performance is even better than the results obtained by manually selecting between the results obtained by the TDNN and HORNN for each meeting. The best c-vector system with overall \%SERs of 11.3 and 10.6 on the original and manual segmentation outperform those given by the TDNN and HORNN d-vector systems by a considerable margin of 21\% relative.

\subsection{AMI and RT05 Diarisation Results}
In this section, the full speaker diarisation pipeline shown in Fig. \ref{fig:pipeline} is used 
%to produce the automatic segmentation 
to process both AMI and RT05. It takes each entire audio stream as the input, and outputs the speech segments along with the relevant speaker labels. 
%speech segments with speaker labels attached.

\subsubsection{Window-level clustering based on VAD output}
The segments obtained using VAD were evaluated against both the original reference and the modified reference, and the results are shown in Table \ref{tab:vad}.
\begin{table}[ht]
    \centering
    \caption{AMI VAD results scored using the original and modified reference files, which are presented in the form of \%MS | \%FA.}
    \begin{tabular}{ccc}
    \toprule
         Dataset   & Original Reference & Modified Reference \\
            \midrule
        Dev & 3.6 | 1.7 & 1.2 | 4.0 \\
        Eval &  4.7 | 2.1 & 1.3 | 3.6 \\
        \bottomrule
    \end{tabular}
    \label{tab:vad}
\end{table}
The minimum non-speech segment duration is used as a hyper-parameter during VAD to determine the segmentation. If the duration of the non-speech segment in between of two speech segments is shorter than the threshold, the non-speech segment is regarded as an intra-utterance silence and will be merged into one segment together with its surrounding speech segments. To achieve a good balance between the MS and FA obtained using the original reference, the threshold needed to be set to 1.2 seconds. The best threshold found for the modified reference is 0.2 seconds, which is a more reasonable maximum length of the intra-utterance silence showing the value of the modified reference.
Therefore, in this section, only the results obtained using the modified reference are considered.

%Afterwards, t
The speech segments obtained using only the VAD stage were then used to perform window-level clustering, which is more directly influenced by the c-vector performance compared to the results obtained with an extra CPD stage. 
%reflects the performance of c-vectors without relying on the quality of CPD segments. 
%The window-level clustering results are shown in Table \ref{tab:auto1}, where SERs can be obtained by subtracting the corresponding MSs and FAs in Table \ref{tab:vad} if needed. 
The window-level speaker clustering results are shown in Table \ref{tab:auto1}, which leads to conclusions similar to those obtained based on the manual segmentation results. The DER results which are shown in Table~\ref{tab:auto2} can be obtained by adding each SER to the corresponding MS and FA in Table \ref{tab:vad}.
%The  conclusions to  manual segments can be drawn, except that bi-linear pooling combination performs the best with both original and modified references.

\begin{table}%[width=2.0\linewidth, pos=t]
    % \centering
    \caption{AMI \%SER results on the automatic segmentation obtained based on the VAD output. Window-level speaker clustering is used for spectral clustering. Scoring is performed against the modified reference.}
    %AMI results in \%SER of different model combination methods on the manual segmentation. ``Original'' and ``Modified'' refer to the results obtained by testing on the original and modified segmentation and scored against the  original and modified reference files accordingly. ``\#Params.'' shows the total number of parameters of each system, and ``Classification Accuracy'' presents the speaker classification accuracy on the held-out validation set. ``Manual section'' is an oracle system obtained by manually switching between the TDNN and HORNN systems according to their \%SERs on Dev or Eval separately.
    \begin{tabular*}{\tblwidth}{Lcc}
    \toprule
    %\multirow{2}{*}{System} & \multicolumn{2}{c}{\%SER} \\
    %\cmidrule(lr){2-3}
    System & Dev & Eval  \\
    \midrule
        $\mathbf{e}_{\text {T}}$ (TDNN) & 17.9 & 18.5 \\
        $\mathbf{e}_{\text {H}}$ (HORNN) & 14.6 & 17.7 \\
        \midrule
        FCFusion$(\mathbf{e}_{\text {T}},\mathbf{e}_{\text {H}})$ &	14.1 & 15.6 \\
        SelfAtt1$(\mathbf{e}_{\text {T}},\mathbf{e}_{\text {H}})$ & \textbf{12.4} & 14.9  \\
        SelfAtt2$(\mathbf{e}_{\text {T}},\mathbf{e}_{\text {H}})$ & 13.5 & 16.6 \\
        GatedAdd$(\mathbf{e}_{\text {T}},\mathbf{e}_{\text {H}})$ & 14.3 & 14.8 \\
        Bilinear$_{\sigma}(\mathbf{e}_{\text {T}},\mathbf{e}_{\text {H}})$ & 12.9 & 13.9 \\
        Bilinear$_{\text{tanh}}(\mathbf{e}_{\text {T}},\mathbf{e}_{\text {H}})$ & 14.0 & {13.1} \\
        \midrule
        Bilinear$_{\sigma}(\text{SelfAtt1}(\mathbf{e}_{\text {T}},\mathbf{e}_{\text {H}}),\mathbf{e}_{\text {H}})$ & 13.6 & \textbf{10.7} \\
        Bilinear$_{\text{tanh}}(\text{SelfAtt1}(\mathbf{e}_{\text {T}},\mathbf{e}_{\text {H}}),\mathbf{e}_{\text {H}})$ & 13.4 & 13.4 \\
    \bottomrule
    \end{tabular*}
    \label{tab:auto1}
\end{table}

\subsubsection{Results with additive combination methods}

\begin{table}[t]
\caption{Comparison among different additive combination methods. \%DER is computed on both AMI Dev and Eval with 34 meetings in total. The numbers of positive and negative signs are the numbers of meetings that the relevant system performs better and worse than FCFusion$(\mathbf{e}_{\text {T}},\mathbf{e}_{\text {H}})$. Detailed meeting level results can be found in Table \ref{tab:meeting1} in the Appendix.}
    \centering
    \begin{tabular}{lccc}
    \toprule
        System & Overall \%DER & Sign + / - & $p$-value \\
        \midrule
        FCFusion$(\mathbf{e}_{\text {T}},\mathbf{e}_{\text {H}})$ & 19.9\% &  N/A & N/A \\
        SelfAtt1$(\mathbf{e}_{\text {T}},\mathbf{e}_{\text {H}})$ & \textbf{18.7}\% & \textbf{22} / \textbf{12} & \textbf{0.032} \\
        GatedAdd$(\mathbf{e}_{\text {T}},\mathbf{e}_{\text {H}})$ & 19.6\% & 21 / 13 & 0.054\\
    \bottomrule
    \end{tabular}
    \label{tab:sign}
\end{table}

%As shown in Fig. \ref{FIG:2dselfatten}, since the ratios of the 2D self-attentive combination fluctuates around 1, it is worthwhile comparing this to the FC fusion in Eqn. \eqref{eq:fcfusion} which effectively performs a direct summation over d-vectors to verify if the fluctuation in combination weights, and the gating mechanism in Eqn. \eqref{eq:gating} brings improvements. 
In this section, a meeting level comparison is performed among the additive combination methods: FCFusion$(\mathbf{e}_{\text {T}},\mathbf{e}_{\text {H}})$, 
SelfAtt1$(\mathbf{e}_{\text {T}},\mathbf{e}_{\text {H}})$, and GatedAdd$(\mathbf{e}_{\text {T}},\mathbf{e}_{\text {H}})$. 
Since SER varies significantly across different meetings, this comparison aims
%to show whether 
to find if the improvement is from only a few meetings that perform a lot better, or it is consistent across the majority of meetings. A sign test is performed on SelfAtt1$(\mathbf{e}_{\text {T}},\mathbf{e}_{\text {H}})$, and GatedAdd$(\mathbf{e}_{\text {T}},\mathbf{e}_{\text {H}})$, both compared to FCFusion$(\mathbf{e}_{\text {T}},\mathbf{e}_{\text {H}})$, with p-values reported. 

% the results produced by the systems with different additive combination methods, FCFusion$(\mathbf{e}_{\text {T}},\mathbf{e}_{\text {H}})$, SelfAtt1$(\mathbf{e}_{\text {T}},\mathbf{e}_{\text {H}})$, 
% and GatedAdd$(\mathbf{e}_{\text {T}},\mathbf{e}_{\text {H}})$ are compared.
% %Therefore, a detailed comparison among these three types of c-vectors is performed. 
% The reason to present such a comparison here since the results of window-level clustering are more directly determined by the performance of the c-vectors than the segment-level clustering.
% %because window-level clustering is not influenced by the averaging effect within segments, hence more directly related to the discriminative ability of c-vectors. 
% Besides the overall SER comparison, a meeting-wise sign test is also performed to show one is indeed better than another, since SER varies significantly across meetings. 

As shown in Table \ref{tab:sign}, SelfAtt1$(\mathbf{e}_{\text {T}},\mathbf{e}_{\text {H}})$ not only achieved a lower overall DER, but also achieved better results over 22 meetings compared with FCFusion$(\mathbf{e}_{\text {T}},\mathbf{e}_{\text {H}})$, which gives a $p$-value better than the significance level of 0.05. Moreover, despite GatedAdd$(\mathbf{e}_{\text {T}},\mathbf{e}_{\text {H}})$ and FCFusion$(\mathbf{e}_{\text {T}},\mathbf{e}_{\text {H}})$ result in similar DERs, the former system achieved better performances over 21 meetings with a $p$-value close to the significance level of 0.05. The significance level confirms the consistent improvement across the majority of meetings.

\subsubsection{Full pipeline results with the CPD stage}
\begin{table}[h]
    \centering
    \caption{The results of different CPD models. Collars of 0.5 second are applied to the ground-truth change points. ``d-vector pretraining'' refers to initialising the TDNN part of the CPD model with a TDNN frame-level d-vector extraction model. ``Joint training'' refers to jointly train the entire CPD model after the d-vector pre-training.}
    \begin{tabular}{lccc}
    \toprule
    System     &  Precision & Recall & F1-score\\
    \midrule
      KL-divergence CPD   & 0.17 & 0.33 & 0.22 \\
      Neural CPD & 0.33 & 0.43 & 0.37 \\
        \ \ + d-vector pre-training & 0.42 & 0.64 & 0.51 \\
        \ \ \ \ + joint training  & \textbf{0.50} & \textbf{0.68} & \textbf{0.57} \\
        \bottomrule
    \end{tabular}
    \label{tab:cpd}
\end{table}
From the full diarisation pipeline shown in Fig.~\ref{fig:pipeline}, segments output from the VAD stage are sent to the CPD stage to generate speaker-homogeneous segments before clustering is performed. The input to the CPD stage covers a 1-second long input window centred at the current frame. If consecutive frames are classified as change points, the segment will be split at the centre frame into two speaker-homogeneous segments. After splitting, segments shorter than 0.3 second will be merged into its surrounding segments to avoid over-segmentation. The values of Precision, Recall and F1-score produced by different CPD systems are given in Table \ref{tab:cpd}. 
The neural CPD is trained from scratch unless explicitly specified. While the neural CPD model proposed in Section~\ref{ssec:vadcpd} trained from scratch clearly outperforms a baseline based on Kullback–Leibler-divergence (KL-divergence) between Gaussian windows \citep{vad2,cmuklcpd}, 
a considerable improvement in the CPD performance is observed, If the TDNN part of the CPD model is pretrained as a frame-level d-vector extraction model.
A further improvement can be found by performing joint training of the entire CPD model. 
%The joint training of the frame-level d-vector initialised CPD system can bring a further improvement.
%The joint-training the entire CPD system brings a further gain in F1-score.

% \begin{table}[h]
%     \centering
%     \caption{The results of different CPD models. Collars of 0.5 second are applied to the ground-truth change points. ``d-vector pretraining'' refers to initialising the TDNN part of the CPD model with a TDNN frame-level d-vector extraction model. ``Joint training'' refers to jointly train the entire CPD model after the d-vector pre-training.}
%     \begin{tabular}{lccc}
%     \toprule
%     System     &  Precision & Recall & F1-score\\
%     \midrule
%       KL-divergence CPD   & 0.17 & 0.33 & 0.22 \\
%       Neural CPD & 0.33 & 0.43 & 0.37 \\
%         \ \ + d-vector pre-training & 0.42 & 0.64 & 0.51 \\
%         \ \ \ \ + joint training  & \textbf{0.50} & \textbf{0.68} & \textbf{0.57} \\
%         \bottomrule
%     \end{tabular}
%     \label{tab:cpd}
% \end{table}

\begin{table*}[width=2.0\linewidth, pos=H]
    % \centering
    \caption{AMI \%DER results of different systems with the automatic segmentation obtained as the VAD output, window-level clustering results, and clustering with the neural CPD output. The results are scored against the modified segmentation.}%Numbers in brackets are scored using original references while numbers outside are scored with modified references.}
    \begin{tabular*}{\tblwidth}{@{} Lcccccc@{} }
    \toprule
     & \multicolumn{3}{c}{Dev} & \multicolumn{3}{c}{Eval} \\
    \cmidrule(lr){2-4} \cmidrule(lr){5-7}
     System & {VAD} & {Window-level} & {Neural}& {VAD} & {Window-level} & {Neural} \\
    & output& clustering& CPD& output& clustering& CPD\\
    \midrule
        $\mathbf{e}_{\text {T}}$ (TDNN)  & 27.7 & 23.1	&	\textbf{23.0} &	23.5 & 23.4 &	\textbf{20.5}\\
        
        $\mathbf{e}_{\text {H}}$ (HORNN) & 22.5 & 19.8 &	\textbf{18.1} & 25.0 & 22.4  &	\textbf{19.7} \\
        \midrule
        FCFusion$(\mathbf{e}_{\text {T}},\mathbf{e}_{\text {H}})$ & 22.4 & 19.3 & \textbf{18.5} & 22.8 & 20.5 & \textbf{18.2} \\
        SelfAtt1$(\mathbf{e}_{\text {T}},\mathbf{e}_{\text {H}})$ & 21.7 & 17.6 &	\textbf{16.4} & 23.7 & 19.8 &	\textbf{18.4} \\
        
        SelfAtt2$(\mathbf{e}_{\text {T}},\mathbf{e}_{\text {H}})$ & 22.7 & 18.7 &	\textbf{16.7} & 25.7 &	21.5 &	\textbf{17.7} \\
        
        GatedAdd$(\mathbf{e}_{\text {T}},\mathbf{e}_{\text {H}})$ & 22.3 & 19.5 & \textbf{17.3} & 19.6 & 19.7 & \textbf{16.8} \\
        
        Bilinear$_{\sigma}(\mathbf{e}_{\text {T}},\mathbf{e}_{\text {H}})$  & 23.3 & 18.1 &	\textbf{16.9} & 20.6	& 18.6 &	\textbf{15.6} \\
        
        Bilinear$_{\text{tanh}}(\mathbf{e}_{\text {T}},\mathbf{e}_{\text {H}})$ & 24.1  & 19.2 &	\textbf{18.1} & 21.9 & 18.0 	&	\textbf{15.9}\\
        \midrule
        Bilinear$_{\sigma}(\text{SelfAtt1}(\mathbf{e}_{\text {T}},\mathbf{e}_{\text {H}}),\mathbf{e}_{\text {H}})$ & 21.9 & 18.8 & \textbf{16.4} & 20.9 & 15.6  & \textbf{15.4} \\
        
        Bilinear$_{\text{tanh}}(\text{SelfAtt1}(\mathbf{e}_{\text {T}},\mathbf{e}_{\text {H}}),\mathbf{e}_{\text {H}})$ & 21.6 & 18.6 & \textbf{16.2} & 23.5 & 18.3 & \textbf{17.8} \\
    \bottomrule
    \end{tabular*}
    \label{tab:auto2}
\end{table*}

The results with different automatic segmentation are presented in Table \ref{tab:auto2}. In addition to window-level clustering, another baseline clusters VAD output directly by treating each VAD segment as a single speaker segment. 
The DERs of all systems after CPD improved compared to the window-level clustering results, and the proposed neural CPD method improves performance in all systems on both the Dev and Eval sets. 
Regarding systems with a single d-vector combination method, SelfAtt1$(\mathbf{e}_{\text {T}},\mathbf{e}_{\text {H}})$ achieves the best DER on Dev, with 9\% relative DER reduction compared to the HORNN d-vector system. Bilinear$_{\sigma}(\mathbf{e}_{\text {T}},\mathbf{e}_{\text {H}})$ performs the best on Eval with 21\% relative DER reduction. By using the two  methods together, Bilinear$_{\sigma}(\text{SelfAtt1}(\mathbf{e}_{\text {T}},\mathbf{e}_{\text {H}}),\mathbf{e}_{\text {H}})$ achieves the lowest DERs across the table. %, which are also the lowest DERs on AMI MDM to the best of the authors' knowledge. 
The relative DER reductions over the HORNN d-vector system are 10\% and 22\% on Dev and Eval respectively. For completeness, the influence of incorporating overlapping speech regions during scoring is shown in Table \ref{tab:finalresult}. For the original reference, including overlap when scoring results in a much higher increase in DER compared to the modified reference. This is mainly due to the amount of overlap caused by excessive silence in the original reference being removed in the modified reference.

\begin{table}[t]
\caption{\%DERs on AMI Dev and Eval of Bilinear$_{\sigma}(\text{SelfAtt1}(\mathbf{e}_{\text {T}},\mathbf{e}_{\text {H}}),\mathbf{e}_{\text {H}})$ with the automatic segmentation with neural CPD. The results were scored against both original and modified references, with or without the overlapping regions.}
\begin{tabular}{ccccc}
\toprule
Score with & \multicolumn{2}{c}{Original} &\multicolumn{2}{c}{Modified} \\
\cmidrule(lr){2-3} \cmidrule(lr){4-5}
overlap & {Dev} & {Eval} & {Dev} & {Eval} \\
\midrule
$\times$ & 18.2 & 18.2 & \textbf{16.4} & \textbf{15.4}\\
$\surd$ & 24.9 & 25.4 & 19.4 & 17.8 \\
\bottomrule
\end{tabular}
\label{tab:finalresult}
\end{table}

Furthermore, the proposed methods were evaluated on the NIST RT05 data set and the results are shown in Table \ref{tab:rt05}. The scoring pipeline is the same as before, which excludes overlapped speech segments. Note that the systems trained on AMI were directly used on the meetings in RT05 which were recorded at 5 different sites and 4 of them are not included in the AMI corpus. The number of speakers in each meeting also varies from 4 to 10, which makes the clustering procedure even more challenging.
The systems trained and tuned on AMI were directly applied to the meetings from RT05. The MS and FA together is 2.6\%, and SERs can be obtained by subtracting 2.6\% from corresponding DERs in Table \ref{tab:rt05}. All of the proposed c-vector systems achieved better performance than any individual d-vector system. The best performance 
%on both sets 
was achieved by Bilinear$_{\text{tanh}}(\mathbf{e}_{\text {T}},\mathbf{e}_{\text {H}})$, 
Bilinear$_{\sigma}(\text{SelfAtt1}(\mathbf{e}_{\text {T}},\mathbf{e}_{\text {H}}),\mathbf{e}_{\text {H}})$, and 
Bilinear$_{\text{tanh}}(\text{SelfAtt1}(\mathbf{e}_{\text {T}},\mathbf{e}_{\text {H}}),\mathbf{e}_{\text {H}})$, which each gave 15\% and 12\% relative reductions in SERs and DERs separately compared to the TDNN baseline system. Moreover, the consistent performance gain and similar DER numbers also reflects the robustness of the proposed systems.
% which are the lowest DER and SER results on RT05, to the best of the authors' knowledge. 

\begin{table}[ht]
    \centering
    \caption{RT05 results without scoring the overlapping regions obtained using our full diarisation pipeline with the neural CPD. SERs can be obtained by subtracting MS and FA, altogether 2.6\%, from corresponding DERs in Table \ref{tab:rt05}}
    \begin{tabular}{lcc}
    \toprule
    System     & \%DER \\
    \midrule
      $\mathbf{e}_{\text {T}}$ (TDNN)   & 16.2 \\
      $\mathbf{e}_{\text {H}}$ (HORNN) & 18.5 \\
      \midrule
      FCFusion$(\mathbf{e}_{\text {T}},\mathbf{e}_{\text {H}})$ & 19.2\\
      SelfAtt1$(\mathbf{e}_{\text {T}},\mathbf{e}_{\text {H}})$ & 14.5 \\
      SelfAtt2$(\mathbf{e}_{\text {T}},\mathbf{e}_{\text {H}})$ & 14.7 \\
      GatedAdd$(\mathbf{e}_{\text {T}},\mathbf{e}_{\text {H}})$ & 16.0 \\
      Bilinear$_{\sigma}(\mathbf{e}_{\text {T}},\mathbf{e}_{\text {H}})$ & 14.3 \\
      Bilinear$_{\text{tanh}}(\mathbf{e}_{\text {T}},\mathbf{e}_{\text {H}})$ & \textbf{14.2}\\
      Bilinear$_{\sigma}(\text{SelfAtt1}(\mathbf{e}_{\text {T}},\mathbf{e}_{\text {H}}),\mathbf{e}_{\text {H}})$ & \textbf{14.2}\\
      Bilinear$_{\text{tanh}}(\text{SelfAtt1}(\mathbf{e}_{\text {T}},\mathbf{e}_{\text {H}}),\mathbf{e}_{\text {H}})$ & \textbf{14.2} \\
    \bottomrule
    \end{tabular}
    \label{tab:rt05}
\end{table}

\subsubsection{Full pipeline results using extra training data}

Finally, to show the effectiveness of the proposed combination methods on large-scale training data as suggested in \citep{AMIlarge}, the joint VoxCeleb \citep{VoxCeleb1} and VoxCeleb2 \citep{VoxCeleb2} data were used following the same data preparation pipeline in \citep{FloriansPaper}. The training data contains 2,789 hours of speech and 7,323 speakers in total. A 512-dim TDNN model and a 512-dim HORNN model were trained on the VoxCeleb data, and then were jointly optimised with combination layers on the AMI training data only. The DER results are reported in Table \ref{tab:voxceleb}, with the same VAD and CPD models used in Table \ref{tab:finalresult} and Table \ref{tab:rt05}. SER results can be obtained by subtracting corresponding MS and FA values. 
\begin{table}[ht]
    \centering
    \caption{\%DER on AMI dev, eval and the RT05 evaluation sets without scoring the overlapping regions obtained using the full diarisation pipeline with the neural CPD. TDNN and HORNN systems were trained on the VoxCeleb 1+2 data, and fine-tuned on the AMI training data. Combination parts were only trained on the AMI training data together with individual systems.}
    \begin{tabular}{lccc}
    \toprule
    System     &  Dev & Eval & RT05 \\
    \midrule
      $\mathbf{e}_{\text {T}}$ (TDNN)    &  12.6 & 15.6	& 12.9 \\ %13.7
      $\mathbf{e}_{\text {H}}$ (HORNN) & 13.8 & 16.7 & 13.0 \\ %14.5
      \midrule
      FCFusion$(\mathbf{e}_{\text {T}},\mathbf{e}_{\text {H}})$ & 14.2 & 16.5 & 13.4 \\
      SelfAtt1$(\mathbf{e}_{\text {T}},\mathbf{e}_{\text {H}})$ & 12.4 & 15.1 & 11.3 \\ %12.9
      SelfAtt2$(\mathbf{e}_{\text {T}},\mathbf{e}_{\text {H}})$ & 12.8 & 14.5 & \textbf{11.2}\\ %12.8
      GatedAdd$(\mathbf{e}_{\text {T}},\mathbf{e}_{\text {H}})$ & 12.3 & 14.6 & 11.5\\ % 12.8
      Bilinear$_{\text{tanh}}(\mathbf{e}_{\text {T}},\mathbf{e}_{\text {H}})$ & \textbf{11.6} & 14.1 & 12.1 \\ %12.6
      Bilinear$_{\sigma}(\mathbf{e}_{\text {T}},\mathbf{e}_{\text {H}})$ & 12.1 & 14.2 & 11.6\\ % 12.6
      Bilinear$_{\text{tanh}}(\text{SelfAtt1}(\mathbf{e}_{\text {T}},\mathbf{e}_{\text {H}}),\mathbf{e}_{\text {H}})$ & 12.1 & 13.9 & 11.3\\ % 12.4
      Bilinear$_{\sigma}(\text{SelfAtt1}(\mathbf{e}_{\text {T}},\mathbf{e}_{\text {H}}),\mathbf{e}_{\text {H}})$ & 12.1 & \textbf{13.8} & 11.3 \\ % 12.4
    \bottomrule
    \end{tabular}
    \label{tab:voxceleb}
\end{table}

DERs on individual systems and combined systems in Table \ref{tab:voxceleb} improved compared to the values in Table \ref{tab:auto2} when VoxCeleb and VoxCeleb2 sets are used for training. For example, the best-performing individual system in Table \ref{tab:voxceleb} is the TDNN whose DERs decreased from 23.0\%, 20.5\% and 16.2\% in Table \ref{tab:auto2} to 12.6\%, 15.6\% and 12.9\% in Table \ref{tab:voxceleb} on the AMI dev, eval and RT05 sets respectively. 

Although the entire combination systems were not optimised on the large-scale data, improvements were found using all combination methods. While the Bilinear$_{\text{tanh}}(\mathbf{e}_{\text {T}},\mathbf{e}_{\text {H}})$ system performed particularly well on meeting IB4002, the best system overall was Bilinear$_{\sigma}(\text{SelfAtt1}(\mathbf{e}_{\text {T}},\mathbf{e}_{\text {H}}),\mathbf{e}_{\text {H}})$ system, which gave rise to 7\%, 17\% and 16\% relatiev SER reductions, and 4\%, 12\% and 12\% relative DER reductions on the AMI dev, eval and RT05 sets respectively. 

% Sec. 7: Conclusion
\section{Conclusions}
\label{sec:conclusions}
In this paper, the 2D self-attentive, gated additive, and bilinear pooling based combination methods are proposed to combine window-level d-vectors to obtain more expressive c-vector speaker embeddings. Furthermore, a complete single pass speaker diarisation pipeline with NN-based VAD and CPD components has also been introduced. By combining both feedforward and recurrent d-vector extraction systems, 
%as the d-vector sub-systems to combine, 
improvements in both SER and DER have been found using all of the proposed combination methods when compared with the single system results. 
Our best-performing model produced state-of-the-art speaker clustering and diarisation results by further stacking the 2D self-attentive and bilinear pooling methods, which
achieved 21\% relative SER reductions on AMI with both manual and automatic segmentation, and 15\% relative SER reductions on RT05. SER improvements to all systems were found when VoxCeleb and VoxCeleb2 data was used for training, and 7\%, 17\% and 16\% relative SER reductions were found on the AMI dev, eval and RT05 sets using the best combination method.
%and also a 21\% relative SER reduction using the complete pipeline. 

\appendix
\section{Detailed Meeting-level results}
\label{append1}
The detailed meeting level SER results from Table~\ref{tab:sign} are presented in Table~\ref{tab:meeting1} which gives results of different additive combination methods.
\begin{table}[H]
 \caption{AMI meeting-level \%SER comparison for the three additive combination methods using window-level clustering and the modified reference.}
    \centering
    \begin{tabular}{l|ccc}
    \toprule
        Meeting & FCFusion & SelfAtt1 & GateAdd \\
        \midrule
         IB4001 &	17.2 &	17.2 &	17.0 \\
         IB4002	&23.3	&23.3	& 23.8 \\
         IB4003	&7.3 &	6.7 &	7.6\\
         IB4004	& 22.2 &	19.0 & 18.1\\
         IB4010 &	4.9	&4.5&	4.0\\
         IB4011&	4.3	& 4.5&	3.5\\
         ES2011a&	29.4 &	29.2&	28.5\\
         ES2011b&	31.2&	12.1&	12.3\\
         ES2011c&	32.8	&32.0&	31.8\\
        ES2011d&	26.7	&25.4&	24.7\\
        IS1008a&	10.0	&4.3	&8.8\\
        IS1008b&	2.8	&2.4	&1.5\\
        IS1008c&	5.2	&4.8	&1.3\\
        IS1008d&	4.0	&4.1	&4.4\\
        TS3004a&	30.0	&36.3&	34.6\\
        TS3004b&	9.2	&8.5	&26.3\\
        TS3004c&	11.1&	11.0&	13.4\\
        TS3004d&	27.3&	13.4&	28.7\\
        EN2002a&	34.5&	9.1	&7.3\\
        EN2002b&	14.6&	14.8&	15.0\\
        EN2002c&	5.7	&5.7	&5.2\\
        EN2002d&	15.2&	14.0&	14.8\\
        ES2004a&	20.5&	18.3&	18.5\\
        ES2004b&	5.7	&5.8	&5.6\\
        ES2004c&	22.9&	23.1&	5.5\\
        ES2004d&	34.8&	34.6&	38.7\\
        IS1009a&	33.8&	44.4&	48.3\\
        IS1009b&	5.8	&5.0	&5.1\\
        IS1009c&	3.8	&3.6	&3.7\\
        IS1009d&	17.5&	18.4&	16.8\\
        TS3003a&	49.3&	36.0&	40.6\\
        TS3003b&	10.5&	10.7&	13.3\\
        TS3003c&	13.8&	14.1&	14.1\\
        TS3003d&	18.3&	18.7&	35.5\\
        \bottomrule
    \end{tabular}
    \label{tab:meeting1}
\end{table}
% Appendix sections are coded under \verb+\appendix+.

% \verb+\printcredits+ command is used after appendix sections to list 
% author credit taxonomy contribution roles tagged using \verb+\credit+ 
% in frontmatter.

%\newpage

\printcredits

\bibliographystyle{cas-model2-names}

\bibliography{cas-refs}


% Author biography without author photo.
% Author biography. Author biography. Author biography.
% Author biography. Author biography. Author biography.
% Author biography. Author biography. Author biography.
% Author biography. Author biography. Author biography.
% Author biography. Author biography. Author biography.
% Author biography. Author biography. Author biography.
% Author biography. Author biography. Author biography.
% Author biography. Author biography. Author biography.
% Author biography. Author biography. Author biography.

% Author biography with author photo.
% Author biography. Author biography. Author biography.
% Author biography. Author biography. Author biography.
% Author biography. Author biography. Author biography.
% Author biography. Author biography. Author biography.
% Author biography. Author biography. Author biography.
% Author biography. Author biography. Author biography.
% Author biography. Author biography. Author biography.
% Author biography. Author biography. Author biography.
% Author biography. Author biography. Author biography.

% Author biography with author photo.
% Author biography. Author biography. Author biography.
% Author biography. Author biography. Author biography.
% Author biography. Author biography. Author biography.
% Author biography. Author biography. Author biography.

\end{document}